\documentclass[usegraphicx,usedcolumn,useAMS,usenatbib]{mn2e}
\usepackage{latexsym}
\usepackage{amsfonts}
\usepackage{amsmath}
\usepackage{amssymb}
\usepackage{multirow}

\title[$\gamma$ rays from massive stars interacting with AGN jets]
{Gamma-ray emission from massive stars interacting with AGN jets}

\author[A. T. Araudo et al.]{A. T. Araudo$^{1}$
\thanks{E-mail: a.araudo@crya.unam.mx},
V. Bosch-Ramon$^{2}$, and 
G. E. Romero$^{3,4}$\\
$^{1}$Centro de Radioastronom\'{\i}a y Astrof\'{\i}sica, Universidad Nacional
Aut\'onoma de M\'exico, A.P. 3-72 (Xangari), 58089 Morelia,\\
 Michoac\'an, M\'exico\\
$^{2}$ Departament d'Astronomia i
Meteorologia, Universitat de Barcelona, Mart\'{\i} i Franqu\`es 1,
08028, Barcelona, Spain\\
$^{3}$Instituto Argentino de
Radioastronom\'{\i}a, C.C.5, (1894) Villa Elisa, Buenos Aires,
Argentina\\ 
$^{4}$ Facultad de Ciencias Astron\'omicas y Geof\'{\i}sicas,
Universidad Nacional de La Plata, Paseo del Bosque, 1900 La Plata,
Argentina} 
\begin{document}


\pagerange{\pageref{firstpage}--\pageref{lastpage}} \pubyear{2007}

\maketitle

\label{firstpage}

\begin{abstract} 
Dense populations of stars surround the nuclear
regions of galaxies. In active galactic nuclei, these stars can
interact with the relativistic jets launched by the supermasive black hole.
In this work, we study the interaction of early-type stars with
relativistic jets in active galactic nuclei. A bow-shaped double-shock 
structure is formed as a
consequence of the interaction of the jet and the stellar wind of each 
early-type star. 
Particles can be accelerated up to relativistic
energies in these shocks and emit high-energy radiation. We compute, 
considering
different stellar densities of the galactic core, the gamma-ray
emission produced by non-thermal radiative processes. This 
radiation may be significant in some cases, and its detection might yield
valuable information on the properties of the stellar population in the galaxy
nucleus, as well as on the relativistic jet. This emission is expected 
to be particularly relevant for nearby non-blazar sources. 
\end{abstract}


\section{Introduction}
\label{sec_intro}

Active galactic nuclei (AGN) consist of a supermassive black hole
(SMBH) surrounded by an accretion disc in the center of a
galaxy. Sometimes these objects  present radio emitting jets
originated close to the SMBH \citep{begelman_rev}. These jets may be
very weak or absent in radio-quiet AGN, but in radio-loud sources
bipolar powerful outflows of collimated plasma are ejected from the inner
regions of the accretion disc.  

Radio-loud AGN produce continuum radiation
along  the whole electromagnetic spectrum, from radio to gamma
rays. The thermal emission is radiated by matter heated during the
accretion process \citep{Shakura-Sunyaev, Bisnovatyi-Kogan-77},
whereas the non-thermal radiation is generated by relativistic
particles accelerated in the jets (e.g.  \citealp{Boettcher}). This
non-thermal emission is thought to be of synchrotron and inverse
Compton (IC) origin \citep[e.g.][]{Ghisellini_85}, although hadronic
models have been also considered to explain gamma-ray sources
\cite[e.g.][]{Mannheim_93, Mucke_01, Aha_02, Matias_Cle_Gustavo,
Matias-Gustavo-blazars}.  In addition to continuum radiation, optical
and ultra-violet emission lines are also produced in AGN. Some of
these lines are broad, emitted by clumps of gas moving with velocities
$v_{\rm g}>1000$~km~s$^{-1}$ and located in a small region close to
the SMBH, the so-called broad line region (BLR).  

The presence of material surrounding the jets of AGN makes jet-medium
interactions likely. For instance, the interaction of BLR clouds with
AGN jets was already suggested by \citet{Blandford-Konigl-79}  as a
mechanism for knot formation in the radio galaxy M87. Also, the
gamma-ray production through the interaction of a cloud from the BLR
with the jet was studied  by \citet{Dar-Laor}, 
and more recently by \citet{blr}. In the
latter work, the authors showed that jet-cloud interactions may generate
detectable gamma rays in non-blazar AGN, of transient nature in nearby
low-luminous sources, and steady in the case of powerful objects.
 
In addition to clouds from the BLR, and also from the Narrow Line
Region (more extended and located further away from the nucleus),
stars also surround the central region of AGN. Jet-star interactions have been
historically studied as a possible mechanism of jet mass-loading and
deceleration in the past. In the seminal work of  \citet{Komissarov_94}, the
interaction of low-mass stars with jets was studied to analyze the
mass transfer from the former to the latter in elliptical
galaxies. Komissarov concluded that in low-luminous jets, the
interaction with stars can significantly affect the jet dynamics and
matter composition. In the same direction, \citet{Hubbard_06} analyzed
the mass loading and truncation of the jet by interactions with stars,
also considering the case of an interposed stellar cluster.

The gamma-ray emission generated by the interaction of massive stars
with (blazar type) AGN jets has been studied by
\citet{Bednarek-Protheroe}.  They  focused on the  gamma-ray emission
reprocessed by pair-Compton cascades in the radiation field of the
star, and produced by relativistic  electrons accelerated in the
shocks formed by the interaction of the  stellar wind with the
jet. Recently, \citet{maxim-RG} studied the interaction of Red Giant
(RG) stars with AGN jets, focusing  on the gamma-ray
emission produced by the interaction between the tidally disrupted
atmosphere of a RG with the inner jet 
\citep[see also][]{maxim-RG-blazars, Mitya_jet-star-13}. 

In the present paper we adopt the main idea of
\cite{Bednarek-Protheroe}, i.e. the interaction of massive stars with
AGN jets, although our scenario consists of a population of massive
stars surrounding the jets, and considers jet-star interactions at
different heights ($z$) of the jet. We analyze the dependence with $z$
of the properties of the interaction region (i.e. the shocks in the
jet and the stellar wind), and also the subsequent non-thermal
processes generated at these shocks.  We consider the injection of 
relativistic electrons and protons, the evolution of these populations
of particles by synchrotron and IC radiative processes in the case of leptons,
and proton-proton interactions for hadrons,
as well as escape losses, and finally the production of X- and gamma
rays. We compute the radiation produced at different distances to the SMBH. 
 In addition,
we consider  the particular case of a powerful Wolf-Rayet (WR) star
interacting at 1~pc from the SMBH. 

In the scenario considered here, the emitters are  the flow
downstream of the bow shocks located around the stars. This flow moves
together with the stars at a non-relativistic speed, and thus the
emission will not be relativistically boosted. For this reason the
radiation from jet-star interactions will be mostly important in
misaligned AGN, where the emission produced by other mechanisms in the
jet (e.g. internal shocks; e.g. \citealt{rees78}) is not amplified by
Doppler boosting\footnote{We neglect emission produced in the shocked 
flows far from the star, where there might be boosting.}. 

Misaligned radio-loud AGN represent an increasing
population of gamma-ray sources. The most populated energy band is the
GeV region, in
which {\it Fermi} has already detected at least 11 sources
\citep{abdo10}, a population that is expected to grow in the near
future. Because of this, theoretical models that can predict the level
and spectrum of the gamma-ray emission from these sources are
timely in order to contribute to the analysis and understanding of
future detections. In this context, jet-massive star interactions are
events that can produce detectable gamma-ray emission in AGN
jets. This phenomenon may be important in spiral  galaxies, where the
star formation rate is high. In addition, some elliptical galaxies
after a violent merger or collision processes
\citep[e.g.][]{Lopez-Sanchez_10} are also expected to harbour a large
number of massive stars near the active core. Finally, star formation
may take place in the external regions of accretion discs of AGN
\citep[e.g.][]{hop10}, and thus a population of massive stars might
exist in the galactic core of even  typical elliptical hosts.

This paper is organized as follows: in Section 2 the main
characteristics of the stellar population near the SMBH are
presented. In Sect. 3, our model of jet-star interaction is described.
In Sect.~4 and 5, the acceleration of particles and the associated
emission are studied. Then, in Sects.~6 and 7, the emission produced
by the interaction of a WR and a population of massive stars
is calculated, and our main results are presented. Finally, a
discussion is given in Sect.~8.

\section{Stellar populations in the nucleus of galaxies}
\label{sec_population}

The characteristics of the stellar populations surrounding the SMBH in
AGN depend on the type of host galaxy. Generally, in spiral galaxies
the  star formation rate $\dot M_{\star}$ is rougly constant, reaching values as
large as $\sim 400$~M$_{\odot}$~yr$^{-1}$ \citep{Mor_12}, 
whereas elliptical galaxies contain large amounts of old stars and 
$\dot M_{\star}$ is very low. However, mergers  between
(elliptical) galaxies can lead to renewed nuclear activity and
episodes of stellar formation \citep[e.g.][]{Sanders-Mirabel-96}, and
accretion of matter to the SMBH may be  associated with star
formation in the galactic nuclei.  In these cases 
$\dot M_{\star} \gtrsim 1000$~M$_{\odot}$~yr$^{-1}$ and the process is episodic. 

The number of stars formed per mass ($m$), time ($t$) and volume
($V \propto r^3$) units can be expressed  as  $\psi(m,r,t) = \psi_0(m,r)
\exp(-t/T)$ \citep{Leitherer-Heckman},  where $\psi_0 \equiv
\psi(t=0)$, and $t$ and $T$ are the age of the stellar system  and the
duration of the formation process, respectively.  There are two limit
cases: continuous formation of stars ($t \ll T$)  and starbursts ($t
\gg T$).  In the former case, $t$ and $T$ are the present age and the
total lifetime  of the host galaxy, respectively, and being $t \ll T$,
$\psi$ can be considered $\sim \psi_0$, and the assumption of a
continuous  and constant star formation process is reasonable.  In the
latter case, $t$ and $T$ are the age and duration of the burst,
respectively, and all the stars are formed almost simultaneously,
implying   $\psi(t \gg T) \sim 0$. In the present work we consider
that stellar formation processes take place continuously in the nuclear 
region of the galaxy, and 
the stars are uniformelly distributed around the SMBH. 
The case of a jet interacting with a massive star forming region will be
considered separately in a future paper.

In the present work we assume that $\psi$ is a power-law mass and radius 
distribution: 
\begin{equation}
\psi = K \left(\frac{m}{M_{\odot}}\right)^{-x}
\left(\frac{r}{\rm pc}\right)^{-y}, 
\label{psi_0}
\end{equation}
where $x \sim 2.3$ for the mass range $0.1 \leq
m/M_{\odot} \leq 120$  \citep{Salpeter_55, Kroupa_01},
$y$ is a free parameter that we fix to  1,  and 2, and
$[K] =$~M$_{\odot}^{-1}$~yr$^{-1}$~pc$^{-3}$. 
Massive stars are formed in giant molecular clouds
with mass $M_{\rm c} \sim 10^3 - 10^7$~M$_{\odot}$ and radius 
$R_{\rm c} \sim 10 - 200$~pc. Stars are formed at a distance from the
SMBH larger than the tidal radius
\begin{equation}
\label{r_t}
r_{\rm t} \sim 2 \left(\frac{M_{\rm bh}}{10^7\,M_{\odot}}\right)^{1/3}
\left(\frac{M_{\rm c}}{10^3\,M_{\odot}}\right)^{-1/3}
\left(\frac{R_{\rm c}}{10\,\rm pc}\right)\,{\rm pc}.
\end{equation}
with a star formation rate 
$\dot M_{\star} = \int\int \psi \,m \,{\rm d}m \,{\rm d}V$, i.e.:
\begin{equation}
\label{M_dot_M}
\dot M_{\star} =  K 
\int_{1 \rm pc}^{1 \rm kpc} \left(\frac{r}{\rm pc}\right)^{-y} 4\pi r^2 {\rm d}r 
\int_{0.1 M_{\odot}}^{120 M_{\odot}} \left(\frac{m}{M_{\odot}}\right)^{-x+1} {\rm d}m.
\end{equation}

\begin{table*}
\caption{Different models considered in the present work. The label
assigned to each model is constructed as $M_{\rm bh}-\eta_{\rm accr}-\eta_{\rm j}$.
For instance, model M7-1-0.01 corresponds to a SMBH with mass 
$M_{\rm bh} = 10^7$~M$_{\odot}$~yr$^{-1}$, $\eta_{\rm accr} = 1$, and 
$\eta_{\rm j} = 0.01$.}
\label{models}
\begin{tabular}{l|l|l|l|l|l|l|l}
\hline\hline
$M_{\rm bh}$& $z_0$ &$L_{\rm Edd}$& $\eta_{\rm accr}$& $\dot M_{\star}$& 
$\eta_{\rm j}$& $L_{\rm j}$ & Model\\
$[M_{\odot}]$& [pc] & [erg~s$^{-1}$]& {}& $[M_{\odot}$~yr$^{-1}]$ & 
{}& [erg~s$^{-1}$] & {}\\
\hline
\multirow{6}{*}{$10^{7}$}& \multirow{6}{*}{$5\times10^{-5}$}& 
\multirow{6}{*}{$1.25\times10^{45}$}& 
\multirow{3}{*}{1}  &\multirow{3}{*}{11.85} & 
0.1 & $1.25\times10^{44}$ & M7-1-0.1\\
\cline{6-8}
& & & & &  0.01 &  $1.25\times10^{43}$& M7-1-0.01\\
\cline{6-8}
& & & & &  0.001 & $1.25\times10^{42}$& M7-1-0.001\\
\cline{4-8}
& & & 0.1 & 1.53 & 0.01 &  $1.25\times10^{43}$& M7-0.1-0.01\\
\cline{6-8}
& & & & & 0.001 &  $1.25\times10^{42}$& M7-0.1-0.001\\
\cline{4-8}
& & & 0.01 & 0.2 & 0.001 &  $1.25\times10^{42}$& M7-0.1-0.001\\
\hline
\multirow{6}{*}{$10^{8}$}& \multirow{6}{*}{$5\times10^{-4}$}& 
\multirow{6}{*}{$1.25\times10^{46}$}& 
\multirow{3}{*}{1}  &\multirow{3}{*}{92.24} & 
0.1 & $1.25\times10^{45}$ & M8-1-0.1\\
\cline{6-8}
& & & & &  0.01 &  $1.25\times10^{44}$& M8-1-0.01\\
\cline{6-8}
& & & & &  0.001 & $1.25\times10^{43}$& M8-1-0.001\\
\cline{4-8}
& & & 0.1 & 11.85 & 0.01 &  $1.25\times10^{44}$& M8-0.1-0.01\\
\cline{6-8}
& & & & & 0.001 &  $1.25\times10^{43}$& M8-0.1-0.001\\
\cline{4-8}
& & & 0.01 & 1.53 & 0.001 &  $1.25\times10^{43}$& M8-0.1-0.001\\
\hline
\multirow{6}{*}{$10^{9}$}& \multirow{6}{*}{$5\times10^{-3}$}& 
\multirow{6}{*}{$1.25\times10^{47}$}& 
\multirow{3}{*}{1}  &\multirow{3}{*}{715.98} & 
0.1 & $1.25\times10^{46}$ & M9-1-0.1\\
\cline{6-8}
& & & & &  0.01 &  $1.25\times10^{45}$& M9-1-0.01\\
\cline{6-8}
& & & & &  0.001 & $1.25\times10^{44}$& M9-1-0.001\\
\cline{4-8}
& & & 0.1 & 92.24 & 0.01 &  $1.25\times10^{45}$& M9-0.1-0.01\\
\cline{6-8}
& & & & & 0.001 &  $1.25\times10^{44}$& M9-0.1-0.001\\
\cline{4-8}
& & & 0.01 & 11.85 & 0.001 &  $1.25\times10^{44}$& M9-0.1-0.001\\
\hline\hline
\end{tabular}
\end{table*}

To obtain $K$, we consider the empirical relation obtained by 
\cite{Satypal_05}:
$\dot M_{\star} = 47.86 (\dot M_{\rm bh}/{\rm M_{\odot}\,yr^{-1}})^{0.89}\,{\rm M_{\odot}\,yr^{-1}}$, where $\dot M_{\rm bh}$ is the SMBH accretion rate. 
Considering that the accretion luminosity
$L_{\rm accr} \sim 0.1 \dot M_{\rm bh} c^2$ is a fraction $\eta_{\rm accr}$ of
the Eddington luminosity, i.e.  $L_{\rm accr} = \eta_{\rm accr} L_{\rm Edd}$, where  
$L_{\rm Edd} = 1.2\times10^{45} (M_{\rm bh}/10^7\,M_{\odot})$~erg~s$^{-1}$
and $M_{\rm bh}$ is the mass of the SMBH, it is
possible to write:
\begin{equation}
\label{M_dot}
\dot M_{\star} = 11.85 
\, \eta_{\rm accr}^{0.89}
\left(\frac{M_{\rm bh}}{10^7 M_{\odot}}\right)^{0.89} \,
{\rm M_{\odot}\,yr^{-1}}.
\end{equation}
On Table~\ref{models}, $\dot M_{\star}$ is given for 
$M_{\rm bh} = 10^7, 10^8$, and $10^9$~M$_{\odot}$,  and $\eta_{\rm accr} = 0.01$,
0.1, and 1. Note that for the nine different combinations of $M_{\rm bh}$ and 
$\eta_{\rm accr}$, we obtain only five  different values of $\dot M_{\star}$,
from 0.2 to 716~M$_{\odot}$~yr$^{-1}$.  Finally, equating
Eqs.~(\ref{M_dot_M}) and (\ref{M_dot}), $K$ results 
\begin{equation}
K \sim 
\left\{\begin{array}{ll}
 3.22\times10^{-7} \,\eta_{\rm accr}^{0.89}
\left(\frac{M_{\rm bh}}{10^7 M_{\odot}}\right)^{0.89}, & y = 1\\
 1.6\times10^{-4} \,\eta_{\rm accr}^{0.89}
\left(\frac{M_{\rm bh}}{10^7 M_{\odot}}\right)^{0.89}, & y = 2.
\end{array}\right.
\end{equation}

\begin{figure*}
\includegraphics[angle=270, width=0.49\textwidth]{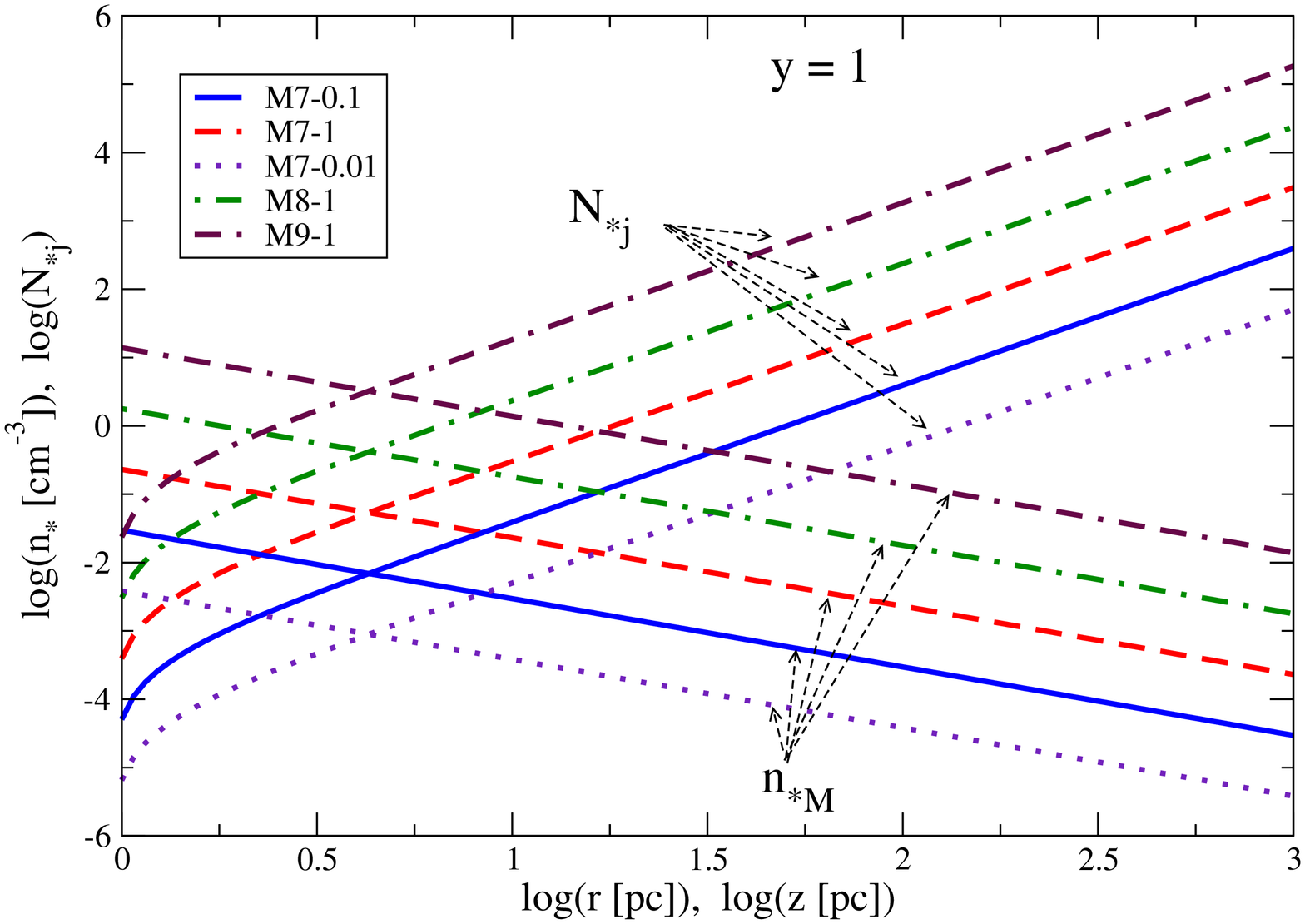}
\includegraphics[angle=270, width=0.49\textwidth]{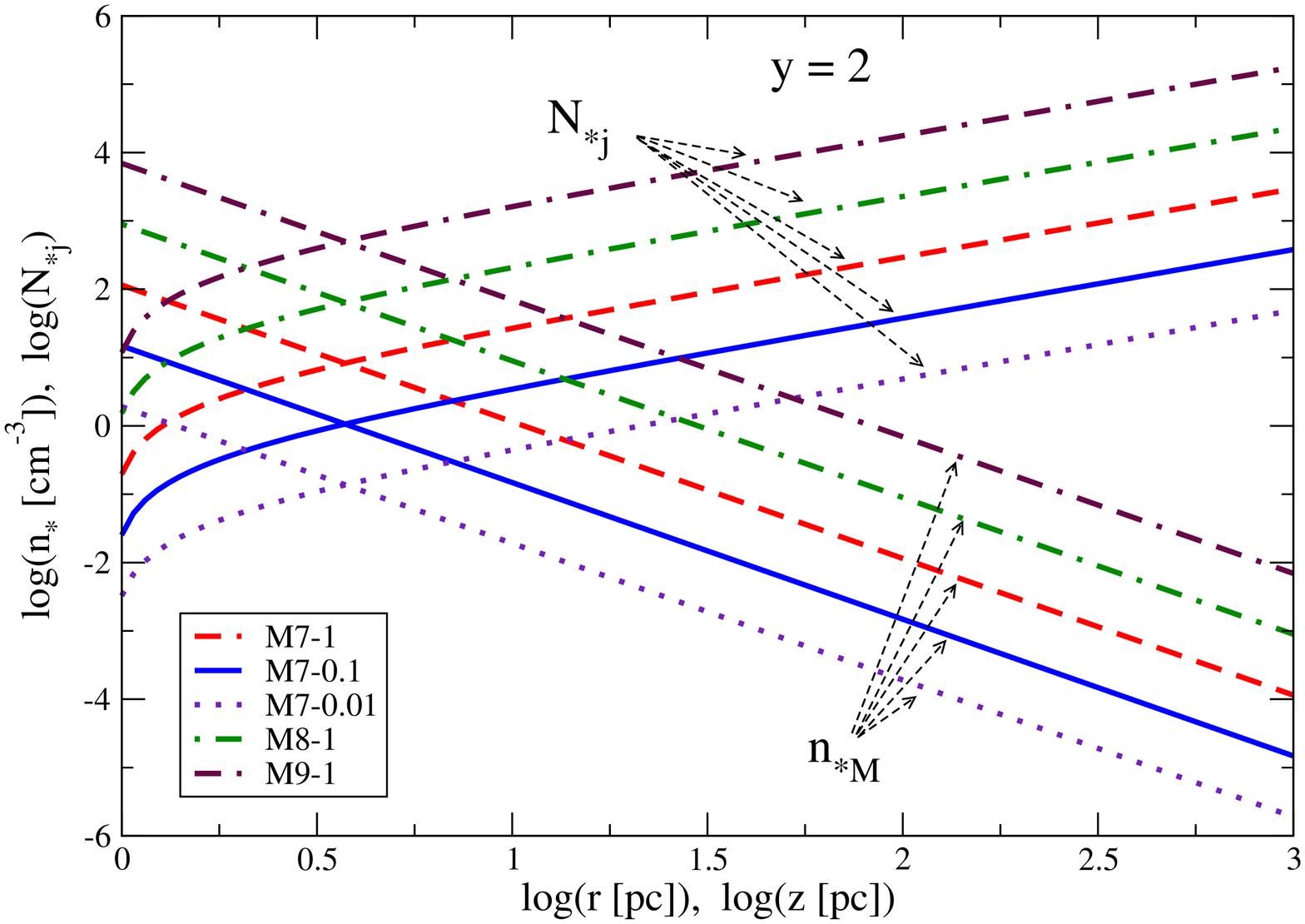}
\caption{Density of massive stars ($n_{\star \rm M}$) and number 
($N_{\star \rm,j}$) of early type stars inside the jet at different values of 
$z$ (that is equivalent to $r$),  and for the case of $y = 1$ (left) and 
$2$ (right). Cases for different combinations of 
$M_{\rm bh}$ and $\eta_{\rm accr}$ are plotted. Other combinations not shown
in the figure provide the same $n_{\star \rm M}$ and $N_{\star \rm j}$ plotted
here.  In the legend box, we did not specify the value of $\eta_{\rm j}$
because the plotted magnitudes are independent of this parameter.}
\label{population_y1}
\end{figure*}

Once a stellar population is injected in the host galaxy, the
new stars will evolve through
collissions with other stars, mass loss by stellar evolution, and 
by stellar disruption through the loss cone (this process will enlarge 
 $M_{\rm BH}$). At the same
time,  stars migrate through the nuclear region forming a central
cluster. Theoretical (e.g \citealt{Murphy_91, Zhao_97}) and 
observational (e.g. \citealt{Schoedel_09}) studies show that stellar systems 
around a SMBH seem to follow a broken power-law 
spatial distribution $n_{\star} = n_{\rm b} (r/r_{\rm b})^{-y_{1,2}}$, where 
$n_{\rm b}$ is the number density at the break radius $r_{\rm b}$, and 
$y_{1}$ and $y_{2}$ are the 
power-law index inside and outside $r_{\rm b}$, respectively. 
The presence of a SMBH produces that the most massive stars are
concentrated around it and, in some cases, a stellar cusp is formed
very close to the event horizon, at $r << r_{\rm b}$, and with a slope 
$\sim -0.5$. 
This region is very small, but the density there is $\sim 10$ times the 
density predicted by $n_{\star} = n_{\rm b} (r/r_{\rm b})^{-y_{1}}$ 
\citep{Murphy_91, Zhao_97}. However, 
in systems with ongoing stellar formation, and low densities,
relaxation timescales
as tidal disruption by the SMBH and collisions between stars can be neglected.
Then, stars of a given mass are accumulated in the galaxy  and, at a time
$t < t_{\rm life}$,  where $t_{\rm life} = a (m/M_{\odot})^{-b}$ 
is the stellar lifetime (in the main sequence), the density of stars is 
\begin{equation}
\label{n_tot}
n_{\star m} = \int_0^{t} \psi(t') \,{\rm d}t'\approx \psi_0\,t
\end{equation}
\citep{Alexander_05_Rev}. For $t>t_{\rm life}$, stars die and the mass
distribution becomes steeper than $-2.3$, following
a  law $n_{\rm \star m}\propto m^{-(2.3+b)}$. In the case of massive stars, 
$t_{\rm life} \sim (m/M_{\odot})^{-1.7}$ and
$\sim 0.1\,(m/M_{\odot})^{-0.8}$~Gyr, for $7 < m/M_{\odot} < 15$
and  $15 < m/M_{\odot} < 60$, respectively \citep{Ekstrom_12}. 
For $m > 60$~M$_{\odot}$, $t_{\rm life}$ is almost constant and around 0.004~Gyr
\citep{Crowther_12}. Then, at
$t \gtrsim t_{\rm life}(8 M_{\odot}) \sim 0.03$~Gyr, the rate of stellar formation
is equal to the rate of stellar death and the system reaches the steady 
state for $m>8\,M_{\odot}$. In such a case, the number density of
massive stars -$n_{\star \rm M}$- keeps the spatial  dependence
of the stellar injection rate, $\psi \propto r^{-y}$, resulting 
\begin{eqnarray}
\label{n_masivas}
n_{\star \rm M} & = \int_{8\,M_{\odot}}^{120\,M_{\odot}}n_{\rm \star m}\,{\rm d}m \sim 
7.13\times10^5\, K \left(\frac{r}{\rm pc}\right)^{-y} \,{\rm pc}^{-3}\nonumber\\
n_{\star \rm M} & \sim
\left\{\begin{array}{ll}
0.23\,\eta_{\rm accr}^{0.89} \left(\frac{M_{\rm bh}}{10^7 M_{\odot}}\right)^{0.89}\left(\frac{r}{\rm pc}\right)^{-1}, & y = 1\\
114.14 \,\eta_{\rm accr}^{0.89} \left(\frac{M_{\rm bh}}{10^7 M_{\odot}}\right)^{0.89}\left(\frac{r}{\rm pc}\right)^{-2}, & y = 2.
\end{array}\right.
\end{eqnarray}
In Figure~\ref{population_y1}, $n_{\rm \star M}$ 
is plotted for the different models described in Table~\ref{models}, and for
the cases of $y = 1$ and 2. 
We can see from the figures that at a distance $\sim 1$~pc from the SMBH
($\sim 10^6\,R_{\rm Schw}$ -$R_{\rm Schw}=2\,G\,M_{\rm bh}/c^2$- for 
$M_{\rm bh} = 10^7\,M_\odot$),  the nominal density of
stars is $\sim 10^4$ and $10$ stars per pc$^3$ for the case of $y = 2$ and
$1$, respectively. This density decreases
abruptly and at a distance $\sim 1$~kpc from the center the density of
massive stars would be much less than 1 star per pc$^3$.  Note that
$n_{\rm \star M}$ depends on $\eta_{\rm accr} \,M_{\rm bh}$,  and
different combinations of  $M_{\rm bh}$ and $\eta_{\rm accr}$ provide the same
value of $n_{\rm \star M}$.

In the next section we calculate the number of massive stars 
that can enter the jet, which is related to the fraction of the volume 
occupied by stars that is intercepted by the jet of the AGN.

\section{Jet-star interaction}

We are interested in the study of the interactions
between massive stars and jets in AGN. In this section, we describe
the main characteristics of the interaction of a massive star with a
relativistic jet.

Jets of AGN are relativistic ($v_{\rm j} \sim c$), with  macroscopic
Lorentz factors $\Gamma \sim 5-10$. The matter composition of the
jets is not well known because depends on a yet incomplete jet
formation theory. Two prescriptions are commonly adopted:  a jet
composed only by $e^{\pm}$ pairs (e.g. Komissarov 1994), and a
lepto-hadronic jet (e.g. Reynoso et al. 2011), i.e.  $n_p = n_e$,
where  $n_p$ and $n_e$ are the number density of
protons and electrons, respectively. In such a case the jet density 
in the laboratory reference frame is
$\rho_{\rm j} = \rho_e + \rho_p = \rho_p[1 + (m_e/m_p)]\sim \rho_p$, being
$m_e$ and $m_p$  the rest mass of electrons and protons,
respectively.  Thus,  we determine the
jet (number) density as  $n_{\rm j} = \rho_{\rm j}/m_p = L_{\rm
j}/[(\Gamma -1)m_pc^2\sigma_{\rm j}v_j]$,  where $L_{\rm j}$ and
$\sigma_{\rm j} = \pi R_{\rm j}^2$ are the jet kinetic luminosity and
section, respectively, and $R_{\rm j}$ its radius.  According to the
current taxonomy of AGN, jets from type I Faranoff-Riley galaxies
(FR~I) are low luminous,  with a kinetic luminosity $L_{\rm j}
<10^{44}$~erg~s$^{-1}$,  whereas FR~II jets have $L_{\rm j} \gtrsim
10^{44}$~erg~s$^{-1}$. The kinetic power of the jet is related with 
$M_{\rm bh}$ through the Eddington luminosity 
as  $L_{\rm j} = \eta_{\rm j}\, L_{\rm Edd}$. 
In FR~II sources, $\eta_{\rm j} \gtrsim 0.02-0.7$ \citep{Ito_FRII},
whereas in FR~I, $\eta_{\rm j} \lesssim 0.01$. In the present work,
we consider $\eta_{\rm j} < \eta_{\rm accr}$ (see Table~\ref{models}). 
For the different models considered, $L_{\rm j}$  goes from 
$1.2\times10^{42}$ to $1.2\times10^{46}$~erg~s$^{-1}$. 

Jets are probably already formed at a distance  
$z_0 \sim 50\,R_{\rm Schw}\approx 5\times10^{-5}(M_{\rm bh}/10^7\,M_{\odot})$~pc 
from the SMBH \citep[e.g.][]{jun99}.
The jet expands as $R_{\rm j} \sim z\tan\theta  \sim \theta z$,
where the half opening angle $\theta$ is $\sim 1^{\circ}-10^{\circ}$.  With
this geometry, the number of massive  stars contained inside the jet
volume $V_{\rm j}$ is  
$N_{\rm \star j}(z) = \int_{1\,{\rm pc}}^{z} n_{\star\rm M}(z'){\rm d}V_{\rm j}$, 
where d$V_{\rm j} = \pi R_{\rm j}^2{\rm d}z'$ 
($z$ is the $r$-coordinate along the jet). This yields: 
\begin{eqnarray}
\label{N_stars}
N_{\rm \star j}(z)& \sim \left\{\begin{array}{ll}
2.89\,\eta_{\rm accr}^{0.89} \left(\frac{M_{\rm bh}}{10^7 M_{\odot}}\right)^{0.89}
\left[\left(\frac{z}{\rm pc}\right)^2 -1\right], & y = 1\\
1.43\times10^3 \,\eta_{\rm accr}^{0.89} \left(\frac{M_{\rm bh}}{10^7 M_{\odot}}\right)^{0.89}
\left[\left(\frac{z}{\rm pc}\right)-1\right], & y = 2.
\end{array}\right.
\end{eqnarray}
At $z\ge z_1 \sim 1.6 \eta_{\rm accr}^{-0.89}(M_{\rm bh}/10^7\,M_{\odot})^{-0.89}$
and $\eta_{\rm accr}^{-0.89}(M_{\rm bh}/10^7\,M_{\odot})^{-0.89}$~pc, for the case
of $y = 1$ and $2$, respectively, there is 
at least one massive star inside the jet at every time (see 
Fig.~\ref{population_y1}). 
For $z$-values such that $N_{\rm \star \rm j}<1$, then $N_{\star \rm j}$ is the 
fraction of time during which there is a star within the jet.

The permanence of stars inside the jet is determined by the jet
crossing time  $t_{\rm j}\sim 2\,R_{\rm j}/v_{\star}\sim 7\times
10^2\,(z/{\rm pc})^{3/2}\,(M_{\rm bh}/10^7\,M_{\odot})^{-1/2}$~yr,
where  $v_{\star} = (2\,G\,M_{\rm bh}/z)^{1/2}\sim 3\times
10^7\,[(M_{\rm bh}/10^7\,M_{\odot}) (z/{\rm pc})]^{-1/2}$~cm~s$^{-1}$ is the velocity at which stars
are moving around the SMBH.  To analyze the interaction of stars with
the jets, we need to know the structure of the shocks formed as a
consequence of the collision of the jet plasma with the stellar
wind. The double bow shock formed around the star (see Komissarov 1994
for a detailed study of the bow-shock structure and stability) depends
not only on the jet properties, but also on the stellar ones, in
particular the stellar wind mass-loss rate and velocity. Main-sequence
massive (OB) stars have typically mass-loss rates  $\dot M_{\rm w}
\sim 10^{-7}-10^{-5}$~M$_{\odot}$~yr$^{-1}$. This mass-loss is
radiatively driven, forming supersonic winds that reach terminal
velocities $v_{\infty} \sim 3000$~km~s$^{-1}$ in the fastest
cases \citep[e.g.][]{Lamers}. The luminosities and surface temperatures of 
OB stars are
$L_{\star} \sim 10^{37}-10^{39}$~erg~s$^{-1}$ and $T_{\star} \sim
3-4\times 10^4$~K, respectively, determining a stellar
radius $R_{\star} = \sqrt{L_{\star}/(4\pi \sigma_{\rm SB} T_{\star}^4)} 
\sim 10$~R$_{\odot}(L_{\star}/3\times10^{38} {\rm erg\,s^{-1}})^{1/2}
(T_{\star}/3\times10^4\,{\rm K})^{-2}$. Here $\sigma_{\rm SB} \sim 
5.67\times10^{-5}$~erg~cm$^{-2}$~K$^{-4}$ is the Stefan-Boltzmann constant. 
In the present work we fix the
stellar and jet parameters to the values listed in
Table~\ref{t_parameters}. 

\begin{table}
\caption{Jet and stellar parameters considered in this work.}
\label{t_parameters}
\begin{tabular}{@{}ll@{}}
\hline\hline
Description & Value\\
\hline
Stellar mass distribution & $m = 8-120$~M$_{\odot}$\\
Stellar mass-loss rate & $\dot M_{\rm w} = 10^{-6}$~M$_{\odot}$~yr$^{-1}$\\
Stellar wind terminal velocity & $v_{\infty} = 2000$~km~s$^{-1}$\\
Stellar luminosity & $L_{\star} = 3\times 10^{38}$~erg~s$^{-1}$\\
Surface temperature & $T_{\star} = 3\times10^4$~K\\
Stellar surface magnetic field & $B_{\rm s}=10$~G\\
\hline
Accretion luminosity & $L_{\rm accr} = \eta_{\rm accr}\,L_{\rm Edd}$\\
Jet kinetic luminosity & $L_{\rm j0} = \eta_{\rm j}\,L_{\rm Edd}$\\
Jet velocity & $v_{\rm j}\approx c$\\
Jet Lorentz factor & $\Gamma_0 = 10$\\
Jet half opening angle & $\theta =5^{\circ}$\\
Jet base & $z_0=50\,R_{\rm Schw}$\\
\hline 
\hline 
\end{tabular}
\end{table}

\subsection{The jet/stellar wind interaction} 
\label{clumps} 

\begin{figure}
\includegraphics[angle=0, width=0.45\textwidth]{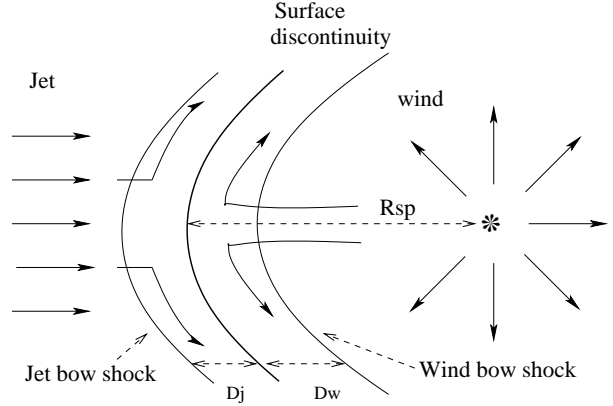}
\caption{Sketch of the double bow-shock configuration formed by the interaction
between the jet plasma and the stellar wind. Jet and wind shocked regions are 
separated by a contact discontinuity, and the shocked matter flows downstream, 
away from the shock apex. $D_{\rm j}$ and $D_{\rm w}$ are the size/thickness of 
the jet and wind bow-shock downstream regions.}
\label{sketch}
\end{figure}

When the jet interacts with the wind of the star, a double bow shock is 
formed, as shown in Fig.~\ref{sketch}.
The stagnation point (SP) is located at a distance $R_{\rm sp}$
from the stellar centre, where the (shocked) wind and jet ram pressures are 
equal.
From $\rho_{\rm w}\,v_{\rm w}^2 = \rho_{\rm j}\Gamma_0\beta_{\rm j}^2 c^2$,
where $\rho_{\rm w}=\dot M_{\rm w}/(4\pi R_{\rm sp}^2 v_{\rm w})$ and 
$\rho_{\rm j}$ 
are the wind and jet densities (in the laboratory reference frame), 
respectively.  Equality of ram pressures yields:
\begin{eqnarray}
\label{R_sp_eq}
\frac{R_{\rm sp,0}}{R_{\rm j}} &\sim&  
10^{-2}\left(\frac{\dot M_{\rm w}}{10^{-6}\,{\rm M_{\odot}\,yr^{-1}}}\right)^{1/2}\times\nonumber\\
{}&\times&\left(\frac{v_{\infty}}{2000\,{\rm km\,s^{-1}}}\right)^{1/2}
\left(\frac{L_{\rm j0}}{10^{42}\,{\rm erg\,s^{-1}}}\right)^{-1/2},
\end{eqnarray}
where we have approximated the wind velocity $v_{\rm w}$ by   $\sim
v_{\infty}$.\footnote{Considering that the wind velocity is described
by a $\beta$-law, i.e. $v_{\rm w} =
v_{\infty}(1-R_{\star}/R)^{\beta}$, where $\beta \sim 0.8-1$, at
distances $R \gtrsim 2\,R_{\star}$, the  approximation $v_{\rm w} \sim
v_{\infty}$ is reasonable.} Note that $R_{\rm sp,0}$ depends on $L_{\rm j0}$,
and then only five combinations of the values of $M_{\rm bh}$ and $\eta_{\rm j}$ 
given in Table~\ref{models} provides different values of $L_{\rm j0}$
(and $R_{\rm sp,0}$).

For the stellar parameters given on
Table~\ref{t_parameters},  $R_{\rm sp}$ will be larger than $R_{\star}$ at  
$z \ge z_{\star\rm w} = 8\,(L_{\rm j0}/10^{42}\,{\rm
erg\,s^{-1}})^{1/2}\,z_0$. Even if the stellar wind were very weak, the jet pressure  might be still balanced
by wind magnetic pressure. For a wind with a surface magnetic field
$B_{\rm s} = 10$~G, this can occur at  $z \gtrsim z_{\star\rm B} =
100\, z_0\, (B_{\rm s}/10\,{\rm G})^{-1}(L_{\rm j0}/10^{42}\,{\rm
erg\,s^{-1}})^{1/2}$; a magnetic field as high as  $10^4$~G would
be required to balance the ram pressure of the jet near its base.   
At $z < z_{\star} \equiv \min\{z_{\star\rm w}, z_{\star\rm B}\}$, the jet flow
will directly collide with the stellar surface and 
$R_{\rm sp,0} \sim R_{\star}$. Either in the case   
the jet ram pressure is balanced by the magnetic field, or by the stellar
surface, a shock can still form in the jet. On the other hand, no
shock will form in the wind.  Note that interactions at $z<z_{\star}$
will be very rare, since $z_{\star} < z_1$ for 
$0.014\,\eta_{\rm j}^{0.5} \eta_{\rm accr}^{0.89} 
(M_{\rm bh}/10^7\,M_{\odot})^{2.39} \lesssim 1$ for both values of $y$.
(To obtain this limit we have considered that $z_{\star} = z_{\star\rm w}$,
because $z_{\star} = z_{\star\rm B}$ only in cases with $B_{\rm s} > 125$~G,
which is not common in massive stars.)

The subscript $0$ at $L_{\rm j}$ and $R_{\rm sp}$ in Eq.~(\ref{R_sp_eq}) 
refers
the fact that jet/star interactions affect the jet power along $z$. The
jet transfers a fraction $\sim (R_{\rm sp}/R_{\rm j})^2$ of its
kinetic energy to the bow shock that is formed around the star, and
therefore $L_{\rm j}$ decreases with $z$. To evaluate this decrease of
$L_{\rm j}$ caused by jet interactions with the stars, we adopt an exponential
dilution factor of the jet kinetic luminosity: $L_{\rm j}(z) = L_{\rm
j0}\exp(-\tau)$, where $\tau$  accounts for the  energy lost 
by all jet-star interactions up to $z$:
\begin{equation}
\label{tau}
\tau(z) = \int_{z_0}^z \left(\frac{\sigma_{\rm sp}}{\sigma_{\rm j}}\right)
\,n_{\star}(z^{\prime})\, \sigma_{\rm j}\,{\rm d}z' =  
\int_{z_0}^z \pi R_{\rm sp}^2 \,n_{\star {\rm M}}(z')\,{\rm d}z',
\end{equation}
where $\sigma_{\rm sp} = \pi R_{\rm sp}^2$ is the bow-shock section. 
Taking this into account,
$R_{\rm sp}$ can be expressed with the following integral equation:
\begin{equation}
\label{trascendente}
R_{\rm sp} = R_{\rm sp,0}(z)\exp\left[\int_{z_0}^z \frac{\pi}{2}\, n_{\star {\rm M}}(z')\,
R_{\rm sp}^2(z'){\rm d}z'\right]\,,
\end{equation}
whose solution is:
\begin{eqnarray}
\label{trascendente2}
R_{\rm sp}(z) & = & \frac{R_{\rm sp,0}}{\sqrt{1- \left(\frac{R_{\rm sp,0}}{z}\right)^2\pi\int_{z_0}^z n_{\star {\rm M}}(z') z'^2 {\rm d}z'}} \nonumber\\
{}&=& \frac{R_{\rm sp,0}}
{\sqrt{1- \left(\frac{R_{\rm sp,0}}{R_{\rm j}}\right)^2 \,N_{\rm \star j}(z)}}\,.
\end{eqnarray}
We have considered here only  the impact of early-type stars because of the
high power of their winds .

\begin{figure}
\includegraphics[angle=270, width=0.49\textwidth]{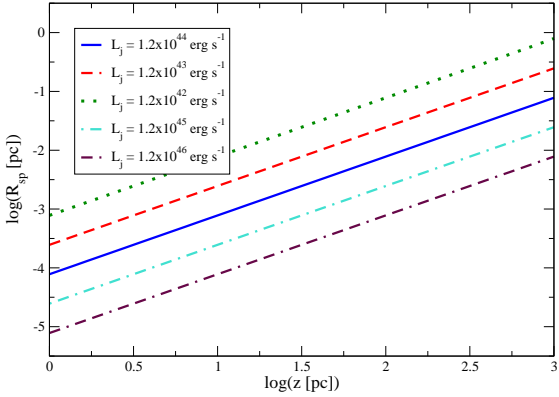}
\caption{Location of the stagnation point $R_{\rm sp}$.
Five values of the jet kinetic luminosity have been  considered: from 
$L_{\rm j} = 1.2\times10^{42}$ to $1.2\times10^{46}$~erg~s$^{-1}$
(see Table~\ref{models}). Note  that at $z \le 1$~kpc, the exponential 
increase of $R_{\rm sp}$ is not present, given $R_{\rm sp} \sim R_{\rm sp,0}$.}
\label{Stag_point}
\end{figure}

Figure~\ref{Stag_point} shows the $z$-dependence of $R_{\rm sp}$ for the 
different cases studied here. In particular, $R_{\rm sp}$ is plotted for
different values of  $L_{\rm j0}$,  from $1.2\times10^{42}$ to 
$1.2\times10^{46}$~erg~s$^{-1}$, and adopting the
parameters of jets and massive stars listed in
Table~\ref{t_parameters}. As shown in the figure, $R_{\rm sp}\sim
R_{\rm sp,0}$ along the whole jet considered here (i.e. up to $z = 1$~kpc). 
For this reason, only the cases with different values of $L_{\rm j0}$ are
plotted in Fig.~\ref{Stag_point}.
The value of $z$ at which
$R_{\rm sp}$ starts to be significantly larger than $R_{\rm sp,0}$ is
related to the condition $\tau > 1$.  At $z < 1$~kpc,
this condition is not fullfiled for any case considered here,
 as is shown in Fig.~\ref{Stag_point}. 
Considering  $R_{\rm sp} = R_{\rm sp,0}$ in Eq.~(\ref{tau}), we  obtain 
an upper limit on the
value $z_2$  at which $\tau = 1$ (i.e. $(R_{\rm sp,0}/R_{\rm j})^2
\,N_{\rm \star j}(z) = 1$). This yields 
\begin{eqnarray}
\label{z2}
\frac{z_2}{\rm kpc} & \sim \left\{\begin{array}{ll}
2\eta_{\rm j}^{0.5}\eta_{\rm accr}^{-0.45} (M_{\rm bh}/10^{7}\,M_{\odot})^{0.06}, 
& y = 1\\
8.4\eta_{\rm j} \eta_{\rm accr}^{-0.89} (M_{\rm bh}/10^{7}\,M_{\odot})^{0.11}, 
& y = 2.
\end{array}\right.
\end{eqnarray}

Since we neglect flow reacceleration downstream the bow shock, or
shading of shocks by other shocks further upstream, when the energy
rate crossing all the shocks reaches $\sim L_{\rm j}$ (i.e. $\tau =
1$),  the jet is completely stopped.  
When $\tau>1$, the approximation of a constant
$L_{\rm j}$ is not valid any more. However, this occurs at $z > 1$~kpc 
on our models.

With the reduction of $L_{\rm j}$ by jet-star interactions, the jet
velocity will decrease. For a cold jet $L_{\rm j}=\dot{M}_{\rm j} (\Gamma -
1) c^2$ and considering $\dot{M}_{\rm j}$ constant,  the Lorentz
factor results $\Gamma=\Gamma_0 \exp(-\tau) + 1$. However, the
assumption of constant $\dot M_{\rm j}$ is not strictly correct. The
entrainment of cold material from the stellar  wind will also
contribute to the deceleration of the jet bulk motion. In the surface
discontinuity a mixing layer will develop, and the shocked jet and
wind matter can mix. This mixing is produced by turbulent motions in
the bow shock tail, likely triggered by Rayleigh-Taylor (RT) and
Kelvin-Helmholtz (KH) instabilities. Under effective mixing,
$\dot{M}_{\rm j}(z) = \dot M_{\rm j}^0 + \dot M_{\star}(z)$, where
$\dot M_{\rm j}^0$ is the initial rate of jet mass and $\dot
M_{\star}(z)\sim N_{\rm \star,j}(z)\,\dot{M}_{\rm w}$.  This effect
has been analyzed by \citet{Komissarov_94} for the case of  low-mass
stars (typical $\dot M_{\rm w}\sim 10^{-12}$~M$_{\odot}$~yr$^{-1}$)
interacting  with jets, concluding that  mixing by KH instabilities is
an important mechanism of mass loading in FR~I galaxies. In the next
subsection we show, through a simple analysis of timescales, that KH
instabilities are also important in the case of massive stars
interacting with jets.  

\subsection{Dynamical timescales}
\label{sec_timescales}
  
We are interested in the  bow shocks  generated around
the stars as places for acceleration of particles and production of
non-thermal emission. For this reason, even when we will not study the
dynamics of these bow shocks, we will estimate the time during which 
stars can be inside the jet as obstacles, and the evolution and
interplay of the shocked flows.

The time required by the star and its wind to penetrate into the jet
is $t_{\rm p}\sim 2\,R_{\rm sp}/v_{\star}\sim 5.6\times
10^2(z/z_0)^{3/2}$~s. In addition to $t_{\rm p}$ and $t_{\rm j}$, 
there are also
hydrodynamical instabilities produced by the jet interaction that
affect the shocked flows, triggering their disruption and mixing. The
timescale for full development of the two bow-shock structures is
roughly $t_{\rm bs}\sim R_{\rm sp}/c_{\rm sw}$, where $c_{\rm sw}$ is
the sound speed in the wind shock, $c_{\rm sw} \sim v_{\rm w}$. This is also the
timescale on which RT and KH instabilities will lead to irregularities
in the contact discontuinity of size $\sim R_{\rm sp}$ \citep[see,
e.g.,][and references therein, for a derivation of these
timescales]{jet-clump}; RT mainly acting in the region around the apex of the
contact discontinuity, and KH in the outflowing tail, further downstream. For
effective disruption of the two shocked flows, and their acceleration
by the jet thrust and eventual mixing, a time of the order of few
times $t_{\rm bs}$ is needed, which yields a length for the mixing
tail of about few times $t_{\rm bs}\,v_{\rm j}\sim R_{\rm
sp}\,\chi^{1/2}$, where $\chi \equiv v_{\rm j}/v_{\rm w}\approx
\sqrt{\rho_{\rm w}/\Gamma\rho_{\rm j}}$. If the ratio $R_{\rm
j}/R_{\rm sp}$ is of the order of or larger than $\chi$, then jet
dilution with $z$ will not have a relevant impact on the
process. Otherwise, jet dilution will likely weaken the instability
growth on the largest tail scales, slowing down mixing.  Effective
mixing also requires that $t_{\rm bs}<t_{\rm j}$, since otherwise the
interaction structure will not fully develop. Given the values of
$R_{\rm sp}$, $R_{\rm j}$, $v_{\star}$ and $v_{\rm w}$ considered in
this work, the mixing conditions seem to be fulfilled, and larger
$M_{\rm bh}$-values (implying larger $v_\star$) should not have a
significant impact. For simplicity, we have kept the reasoning
at a basic level. For a more accurate and detailed description of tail
disruption within relativistic jets, we refer to
\cite{Blandford-Konigl-79} and \cite{Komissarov_94}.


\section{Non-thermal particles}
\label{part-acc}

In addition to the dynamical processes described above, non-thermal
particles can be generated in jet-star interactions. In the bow
shocks, particles can be accelerated up to relativistic energies
through a Fermi-like type~I acceleration mechanism. The size of the
jet and wind shocked regions, $D_{\rm j}$ and $D_{\rm w}$,
respectively, are determined considering the conservation of the rate
of the number particle density. Using relativistic and non-relativistic
Rankine-Hugoniot relations\footnote{We have considered the Rankine-Hugoniot 
relations obtained for the case of a plane shock.} we obtain 
that $D_{\rm j} \sim D_{\rm w}\sim 0.3R_{\rm sp}$.

Although the jet kinetic luminosity is much larger than the wind luminosity 
($L_{\rm w} = \dot M_{\rm w} v_{\infty}^2/2$) at
the location of $R_{\rm sp}$,
\begin{eqnarray}
\frac{L_{\rm j}}{L_{\rm w}} &=& 5\times10^{5}
\left(\frac{L_{\rm j}}{10^{42}\,{\rm erg~s^{-1}}}\right)
\left(\frac{\dot M_{\rm w}}{10^{-6}\,{\rm M_{\odot}~yr^{-1}}}\right)^{-1}
\times\nonumber\\
{}&{}& \times\left(\frac{v_{\infty}}{2000\,{\rm km~s^{-1}}}\right)^{-2},
\end{eqnarray}
the available luminosity in the jet and wind bow shocks, $L_{\rm jbs}$ and
$L_{\rm wbs}$, respectively, are not so different:
$L_{\rm jbs}/L_{\rm wbs}\sim 75\,(v_{\infty}/2000\,{\rm km\,s^{-1}})^{-1}$.

A fraction $\eta_{\rm nt}$ of these luminosities is transferred to
particles accelerated in each shock, implying a non-thermal luminosity
in the jet  $L_{\rm ntj} = \eta_{\rm nt}\,L_{\rm jbs}$, and in the wind
$L_{\rm ntw} = \eta_{\rm nt}\,L_{\rm wbs}$.  The fraction
$\eta_{\rm nt}$  is a free parameter of the present model. We assume
that the populations of accelerated electrons and protons have the
same luminosity, and we fix $\eta_{\rm nt} = 0.1$ both in the jet and
in the wind bow shocks. We note that the radiation luminosity scales
simply as $\propto \eta_{\rm nt}$.  

Relativistic particles are assumed to be injected in the downstream
region of the bow shocks following a power-law energy distribution:
$Q_{e,p}\equiv K_{e,p}\,E_{e,p}^{-2.1}\,\exp(-E_{e,p}/E_{e,p}^{\rm
max})$,  where $E_{e,p}^{\rm max}$  is the maximum energy achieved by
particles, and $e$ and $p$ stands for electrons and protons,
respectively. A power-law index $\sim -2$ is usual for Fermi I-type
acceleration mechanisms, and the normalization constants $K_{e,p}$ are
determined through $L_{\rm nt} = \int Q_{e,p}\, E_{e,p}\,{\rm
d}E_{e,p}$.

As a consequence of radiative and escape losses, the injected particles evolve
until they reach the steady state, with characteristic timescales
$t_{\rm adv,j} \sim 3\,R_{\rm sp}/c$ and $t_{\rm adv,w} \sim
4\,R_{\rm sp}/v_{\infty}$, i.e. the advection escape times in the
downstream regions of the jet and the wind bow shocks, respectively.
In this work we consider that the emitting regions are uniform,
i.e. we adopt a one-zone model for the accelerator/emitter.  Under
this condition, we solve the following equation to derive the energy
distribution of relativistic electrons and protons, $N_{e,p}$
\citep{Ginzburg}: 
\begin{equation}
\label{evolution} 
\frac{N_{e,p}}{t_{\rm esc}}-\frac{{\rm d}}{{\rm d} E_{e,p}}(\dot E_{e,p} N_{e,p}) =Q_{e,p}\,,
\end{equation}
where $t_{\rm esc} = \min\{t_{\rm adv}, t_{\rm diff}\}$.
The diffusion timescale is
$t_{\rm diff} \sim D_{\rm j,w}^2 q B_{\rm jbs,wbs}/(E_{e,p}\,c)$ in
the Bohm regime, where $B_{\rm jbs}$ and $B_{\rm wbs}$ are the magnetic fields in
the jet and the stellar wind bow-shock regions, respectively, and $q$ is the 
electron charge.  In addition to diffusion,
particles suffer different relevant radiative losses $\dot E_{e,p}$, 
synchrotron and stellar photon IC upscattering for electrons, and  
proton-proton \citep{kelner_06}. 
All mentioned losses balance the energy gain from acceleration,
$\dot E_{e,p}^{\rm acc}$, when the steady state is achieved. 

\subsection{Particle acceleration and losses in the jet shock}

The fraction  of the jet section that is intercepted by the stellar 
bow shock, $\eta_{\rm j} = \sigma_{\rm sp}/\sigma_{\rm j}$, is 
$\propto L_{\rm j}^{-1}$. Therefore, $L_{\rm jbs} = \eta_{\rm j}\,L_{\rm j}$  
results to be independent of $L_{\rm j}$ and $z$: 
\begin{eqnarray}
L_{\rm jbs} &=& \left(\frac{R_{\rm sp}}{R_{\rm j}}\right)^2 L_{\rm j} \\
{}&\sim  &
10^{38}\left(\frac{\dot M_{\rm w}}{10^{-6}\,{\rm M_{\odot}\,yr^{-1}}}\right)
\left(\frac{v_{\infty}}{2000\,{\rm km\,s^{-1}}}\right)
\,\,{\rm erg\,s^{-1}}\nonumber.
\end{eqnarray}
Note however that for rare cases of stars interacting 
at $z < z_{\star}$, $R_{\rm sp} = R_{\star}$ and $L_{\rm jbs}$ is 
$\propto L_{\rm j}\,z^{-2}$.
The jet bow shock has a velocity $\sim v_{\rm j}$, and
particles are accelerated at this relativistic shock with a rate assumed to be
$\dot E_{e,p}^{\rm acc} \sim 0.01\, q\, B_{\rm jbs}\, c$. We adopt
a modest acceleration efficiency, although for relativistic
shocks, values as high as 
$\dot E_{e,p}^{\rm acc}\sim 0.1\, q\, B_{\rm jbs}\, c$
have been derived (e.g. \citealt{Achterberg}).

Theoretical studies on jet acceleration \citep[e.g.][]{komissarov_07} 
suggest that
near the base of the outflow, the kinetic energy density of the jet, 
$U_{\rm kin} = L_{\rm j}/(\sigma_{\rm j} v_{\rm j})$,
is  smaller than the magnetic energy density 
$U_{\rm mag} = B_{\rm j}^2/8\pi$, where $B_{\rm j}$ is the jet magnetic field. 
However, at 
$z \gtrsim 10^{-3}(M_{\rm bh}/10^7\,M_{\odot})(\theta/5^{\circ})^{-1}$~pc, 
magnetic forces have already accelerated the flow and 
$U_{\rm kin}$ is likely to be dominant. Given that we are interested on
the jet properties at $z \geq 1$~pc, we estimate $B_{\rm j}$ 
assuming that $U_{\rm mag} = \eta_{\rm B} U_{\rm kin}$, with 
$\eta_{\rm B} = 0.3$ (see Fig.~8 of \cite{komissarov_07} for the case of
conical jets). The corresponding magnetic field is:
\begin{equation}
\label{B_equip}
B_{\rm j} \sim 0.34
\left(\frac{\eta_{\rm B}}{0.3}\right)^{1/2}
\left(\frac{L_{\rm j}}{10^{42}\,{\rm erg\,s^{-1}}}\right)^{1/2}
 \left(\frac{\theta}{5^{\circ}}\right)^{-1}
 \left(\frac{z}{\rm pc}\right)^{-1}\,{\rm G}.
\end{equation}
Then, assuming that in the bow shock downstream region is
amplified by the compression of the flow, $B_{\rm jbs}$ results
$\sim 4\,B_{\rm j} \sim 1.4 [(\eta_{\rm B}/0.3)(L_{\rm j}/10^{42}\,{\rm erg\,s^{-1}})]^{1/2} (z/{\rm pc})^{-1}$~G.

The most important radiative losses of relativistic electrons in the
jet bow-shock region are synchrotron and IC scattering  of photons
from the star. At $R_{\rm sp}$, the energy density of photons is
$U_{\star}\approx L_{\star}/(4\pi R_{\rm sp}^2\,c) \sim
(L_{\rm j}/10^{42}{\rm erg\,s}^{-1})(z/\rm pc)^{-2}$~erg~cm$^{-3}$. 
Considering that these
photons follow a thermal distribution with a maximum at an energy
$E_0\approx 3K_{\rm B}T_{\star} \sim 10 (T_{\star}/3\times 10^4\,{\rm
K})$~eV ($K_{\rm B} = 1.4\times10^{-16}$~erg~K$^{-1}$ is the
Boltzmann constant), at $E_e > (m_e\,c^2)^2/E_0 \sim 50$~GeV, the IC
interaction occurs in the Klein-Nishina (KN) regime.  Photons
from the accretion disc are a less important target for IC
interactions compared with photons from the star, as seen from
the large value of the ratio $U_{\star}/U_{\rm d}\sim 10^2$, for the
wind parameters adopted here and adopting a disc luminosity $\sim
L_{\rm j}$. Electrons can also radiate through relativistic
Bremsstrahlung in interactions with the shocked jet
matter. Nevertheless, densities are so low that relativistic
Bremsstrahlung losses are quite innefficient when compared with
escape, synchrotron or IC scattering.

\begin{figure}
\includegraphics[angle=270, width=0.49\textwidth]{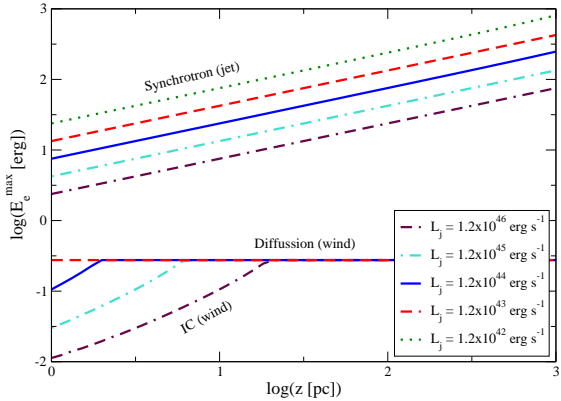}
\caption{Maximum energies of electrons accelerated in the jet (top)
and wind (bottom) bow shocks at different $z$, and for jet kinetic 
luminosities
from $L_{\rm j} = 1.2\times10^{42}$ to  $1.2\times10^{46}$~erg~s$^{-1}$.}
\label{E_max}
\end{figure}

\begin{figure*}
\includegraphics[angle=270, width=0.49\textwidth]{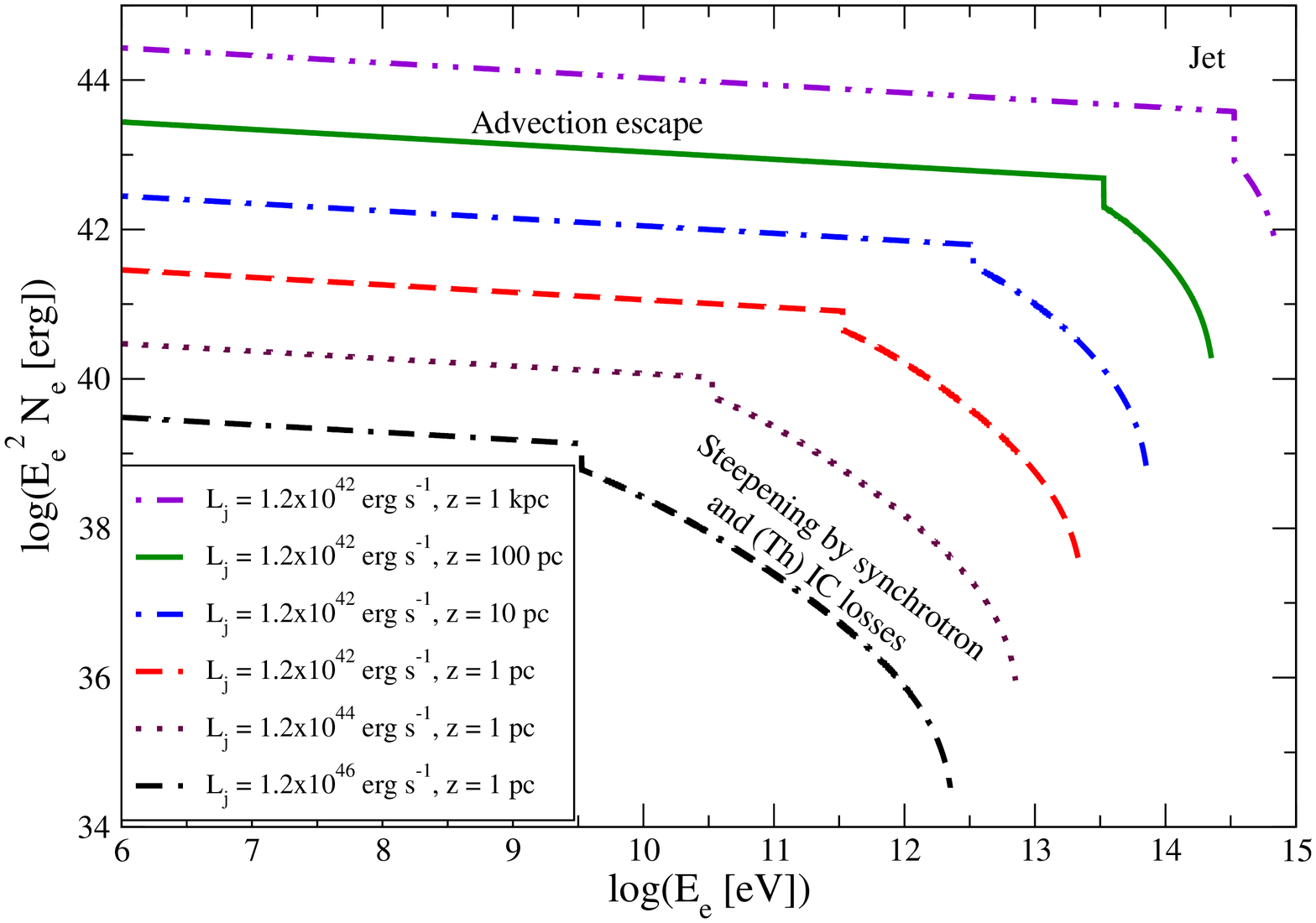}
\includegraphics[angle=270, width=0.49\textwidth]{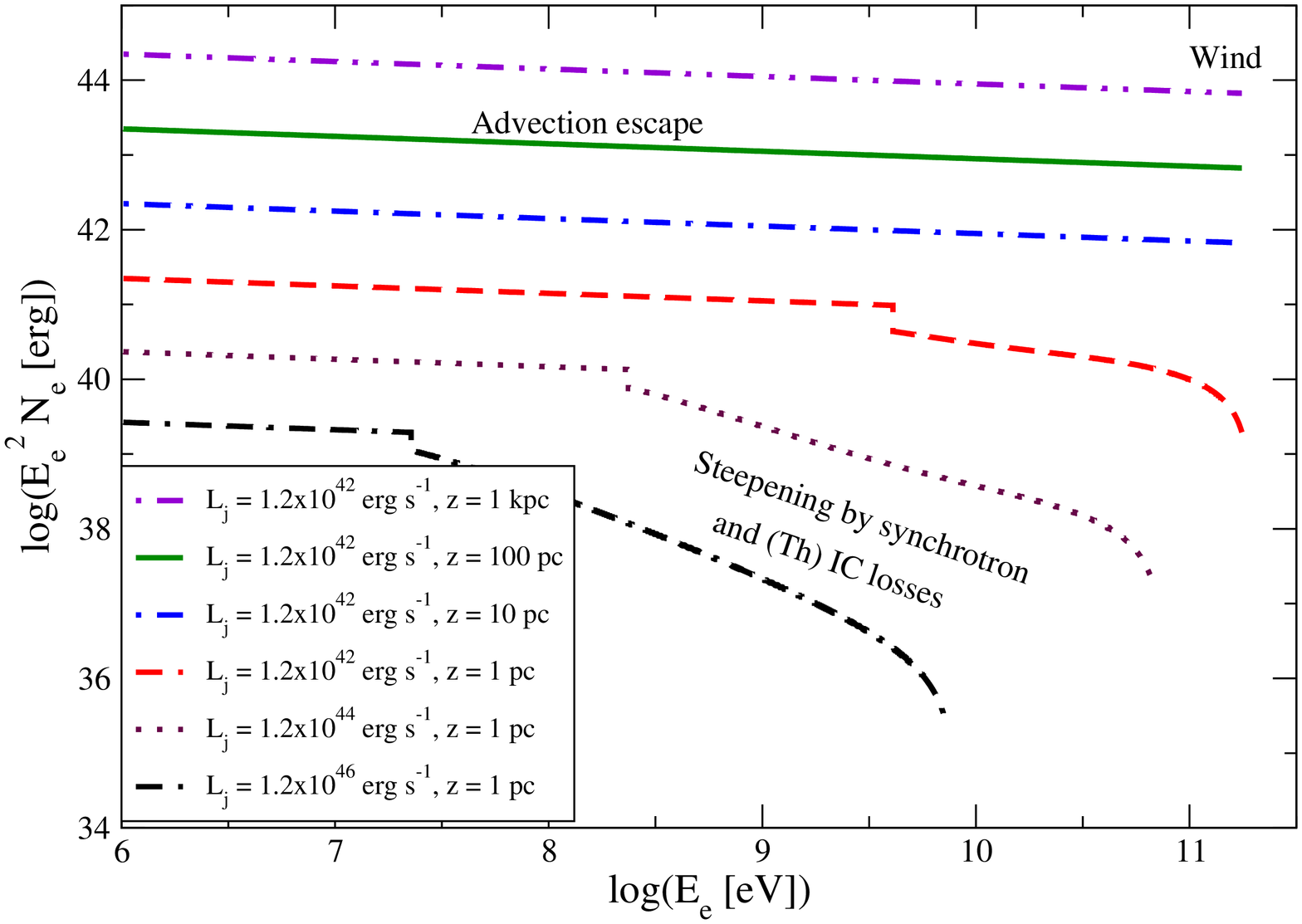}
\caption{Spectral energy distribution of electrons accelerated in the jet
(left) and in the wind (right) bow shock at $z = 1$, 10, 100, and $10^4$~pc), 
 for $L_{\rm j} = 1.2\times10^{42}$, $1.2\times10^{44}$, 
and $1.2\times10^{46}$~erg~s$^{-1}$. The cases with 
$L_{\rm j} = 1.2\times10^{44}$~erg~s$^{-1}$ at $z = 10, 100$, and $10^4$~pc are 
equal to the cases with $L_{\rm j} = 1.2\times10^{42}$~erg~s$^{-1}$ at 
$z = 1, 10$, and $100$~pc, respectively. Also, the cases with 
$L_{\rm j} = 1.2\times10^{46}$~erg~s$^{-1}$ at $z = 10, 100$, and $10^4$~pc are 
equal to the cases with $L_{\rm j} = 1.2\times10^{44}$~erg~s$^{-1}$ at 
$z = 1, 10$, and $100$~pc, respectively.}
\label{espectro_e}
\end{figure*}

The maximum energy achieved by electrons in the jet shock is
determined by  synchrotron losses resulting
\begin{equation}
\frac{E_e^{\rm max}}{\rm TeV} \sim 
20.3 \left(\frac{\eta_{\rm B}}{0.03}\right)^{-1/4}
\left(\frac{z}{{\rm pc}}\right)^{1/2}
\left(\frac{L_{\rm j}}{10^{42}\,{\rm erg\,s^{-1}}}\right)^{-1/4}
\label{e_max_eq}
\end{equation}
In Fig.~\ref{E_max}, $E_e^{\rm max}$ is plotted for different values of 
$L_{\rm j}$.
Taking into account the escape,  synchrotron, and IC losses described
above, we solve Eq.~(\ref{evolution}) obtaining the energy
distribution $N_e$ of relativistic electrons shown in
Fig.~\ref{espectro_e} (left).  
The synchrotron and IC cooling dominate
the high-energy part of the electron energy distribution, and 
at low energies  advective losses are dominant. 
This appears as a 
steepening in $N_e$ from $\propto E^{-2.1}$ to $\propto E^{-3.1}$.

The maximum energy of protons accelerated in the jet  shock is
determined by advection losses, giving $E_p^{\rm max} \sim 2\times10^{4}
(\eta_{\rm B}/0.03)^{-1/2}$~TeV. 
These relativistic protons escape from the jet bow-shock region
advected by shocked matter, without producing significant levels of
radiation. For this reason we do not take into account hadronic
emission from the jet shocked region.  Given that the proton energy is
below the photomeson production threshold with stellar photons as
targets, this process can also be neglected.

\subsection{Particle acceleration and losses in the wind shock} 

Assuming that  the whole wind is shocked, the shock luminosity would be:
\begin{equation}
L_{\rm wbs}\approx 1.3\times10^{36}
\left(\frac{\dot M_{\rm w}}{10^{-6}\,{\rm M_{\odot}\,yr^{-1}}}\right)
 \left(\frac{v_{\infty}}{2000\,{\rm km\,s^{-1}}}\right)^2\,\,{\rm erg\,s^{-1}}.
\end{equation}
Being this shock non-relativistic, with velocity $\sim v_{\infty}$,
particles are accelerated with a rate 
$\dot E_{e,p}^{\rm acc} = (1/2\pi)\,q \,(v_{\infty}/c)^2\,B_{\rm w} \,c $ 
\citep[e.g.][]{dru83}. 

The magnetic field of the wind, $B_{\rm w}$, roughly has a dipolar structure 
close to the star surface, and radial and toroidal components dominate
farther out \citep{usov_92}. 
For simplicity, we will adopt here $B_{\rm wbs} \sim B_{\rm w}$. 
Fixing $B_{\rm s} = 10$~G, $B_{\rm wbs}$ results
$\sim 0.1 B_{\rm jbs}$ at $z > z_{\star}$, and synchrotron cooling will be more
efficient in the jet than in the wind shocked region.
On the other hand, given that the size and
radiation field values are similar, the IC cooling timescale
in the wind shocked region is similar to the one in the jet. The main
difference is in the advection timescale and the maximum energy, given
the much lower shock velocity. 
The lower advection speed implies that the
electron energy distribution steepens at lower energies, implying a
high radiation efficiency. 
The maximum energy of electrons accelerated in the wind is determined
by IC and diffusion losses, providing the values of $E_{e}^{\rm max}$ 
plotted in Fig.~\ref{E_max}.\footnote{We cannot provide an analytical 
expression for $E_{e}^{\rm max}$ in the wind because in the range where it is 
constrained
by IC scattering in the KN regime, the calculation was done through the
Runge-Kutta numerical method.} The lower maximum energy, for the same
non-thermal fraction, also increases the normalization of the energy
distribution. Therefore, although the energetics of the wind shock is
$\sim 100$ times smaller than that in the jet shock, the contribution
of accelerated electrons in the former to the non-thermal output may
be significant. The resulting $N_e$ is shown on Fig.~\ref{espectro_e} (right),
and it is similar to the distribution of
electrons accelerated in the jet, i.e. 
at low values of $z$ $N_e$ is $\propto E_e^{-3.1}$ as a consequence of
IC and synchrotron losses, with a hardening beyond $\sim 10$~GeV.
At larger heights, $N_e \propto E_e^{-2.1}$ all the way up to $E_e^{\rm max}$ 
as a consequence of advection escape losses.

Regarding protons, the large wind particle densities imply that the 
proton-proton cooling channel is more efficient than in the jet shocked 
region, but still it is a minor channel of gamma-ray production compared with 
IC for the same $e$ and $p$ energetics. The proton energy distribution is
dominated by advection losses, which are independent of energy, and
therefore it keeps the injection slope, i.e. $N_p\propto E_p^{-2}$. 
The maximum energy of protons is constrained by diffusion losses, 
giving $E_p^{\rm max} = 0.2(B_{\rm s}/10\,{\rm G})(v_{\rm w}/2000\,{\rm km \,s^{-1}})$~TeV.

\section{Non-thermal emission}
\label{non-thermal}

Once $N_e$  in the jet and wind shocked regions is computed,
we calculate the spetral energy distribution (SED) of the non-thermal
radiation, synchrotron and IC scattering (in Th and KN regimes) 
in the jet and the wind shocked regions,  using the standard fomulae 
(e.g. \citealp{blumenthal_gould}). The energy budget for the emission
produced in the bow shock regions are $\eta_{\rm nt}L_{\rm jbs}$ and 
$\eta_{\rm nt}L_{\rm wbs}$,
which would be an upper limit for the emission luminosity produced both in
the jet and in the wind, respectively.

An important characteristic of the scenario studied in this paper is
that the  emitter is fixed to the star, and being the star moving at a
non-relativistic velocity, the emission produced in the bow shock
regions is not amplified by Doppler boosting.  

At radio wavelengths, the synchrotron self-absorption effect
has been taken into account, although it is only relevant for
interactions very close to the jet base.
At gamma-ray energies, photon-photon absorption due to the presence of
the stellar radiation field can be relevant at certain $z$
\citep[e.g.][]{Bednarek-Protheroe}, but  the internal absorption due
to synchrotron radiation is negligible. Given the typical stellar
photon energy $E_0 \sim 10$~eV, gamma rays beyond $\sim 30$~GeV can be
affected by photon-photon absorption.  However, this process is only
important at small $z$, where $R_{\rm sp}$ is also small.  At $z > 1$~pc 
SEDs shown in Fig.~\ref{sed_zfijo} 
 are not strongly absorbed. Another effect that
should be taken into account at energies beyond 100~GeV is absorption
in the extragalactic background light via pair creation (important only
for sources located well beyond 100~Mpc).

The leptonic emission is indistinguishable if $R_{\rm sp}$ is the
same, regardless the $z$ of interaction and $L_{\rm j}$.  However,
more powerful jets have a transition from radiation to advection
dominated interactions at higher $z$-values, which enhances their
non-thermal luminosity. 
Synchrotron and IC losses are proportional to magnetic (energy) and
radiation  densities, and thus are $\propto z^{-2}$. The increase of the
time during which  particles remain in the emitter, $\propto z$, and the
growth of the number of stars within a jet slice, $\propto z^{0.25}$,
are not enough to balance the loss in radiation efficiency beyond the
$z$ at which radiation cooling is not dominant (at any particle
energy). This implies that there is more emission generated at
relatively small $z$-values.  To illustrate the changes in
the  SED with $z$,  we present in Fig.~\ref{sed_zfijo} the
synchrotron and IC emission produced by the interaction of only one
star with the jet at $z = 1$, 10, 100, and $10^4$~pc, an adopting the
parameters listed on Table~\ref{t_parameters} for the different models 
presented on Table~\ref{models}. (A detailed description of 
Fig.~\ref{sed_zfijo} is given in Sections~\ref{leptonic_j} and 
\ref{hadronic_w}.) In addition to that, we calculate the bolometric 
luminosities achieved by synchrotron and IC emission in the jet 
-$L_{z}^{\rm j}$- and in the wind -$L_{z}^{\rm w}$- by the interaction of only 
one star at different $z$: from 1~pc to 1~kpc. In Fig.~\ref{L_bol}, 
$L_{z}^{\rm j}$ and 
$L_{z}^{\rm w}$  (maroon-solid lines) are shown. (A detailed description of 
this figure is given in Sections~\ref{leptonic_j} and 
\ref{hadronic_w}, and also in Sec.~\ref{Many_stars}.)

\subsection{Leptonic emission from the jet shock}
\label{leptonic_j}

The synchrotron and IC emission from the jet bow shock is presented
in Fig.~\ref{sed_zfijo} (left panel).
As mentioned,  both synchrotron and
IC are more efficient in the inner jet regions, emission at lower
energies getting less efficient (due to advection) at higher
$z$-values. This effect is clearly seen in Figs.~\ref{sed_zfijo} and
\ref{L_bol} (both in the left panel). 

In Fig.~\ref{sed_zfijo} we see different spectral features in the cases of 
$L_{\rm j} = 1.2\times10^{42}$ (top) and $1.2\times10^{46}$~erg~s$^{-1}$ (bottom). 
In the former case, the break energy in the photon spectrum is higher than in 
the latter case. This is very clear in the  synchrotron emission, where
the break energy in the case of $L_{\rm j} = 1.2\times10^{42}$~erg~s$^{-1}$
is about $10^3$ times larger than in the case of   
$L_{\rm j} = 1.2\times10^{46}$~erg~s$^{-1}$.
(Compare  Figs.~\ref{sed_zfijo} and  \ref{espectro_e}.)
Another clear difference is the 
break produced by synchrotron self absorption, being the source optically 
thin at lower energies in the case of  $L_{\rm j} = 1.2\times10^{42}$~erg~s$^{-1}$
than in the case of $L_{\rm j} = 1.2\times10^{46}$~erg~s$^{-1}$.
Photon-photon absorption in the IC spectrum is not relevant in any case.

The total bolometric luminosity produced in the jet, 
$L_{\rm z}^{\rm j} = L_{\rm IC,z}^{\rm j} + L_{\rm synchr,z}^{\rm j}$, where
$L_{\rm IC,z}^{\rm j}$ and $L_{\rm synchr,z}^{\rm j}$ are the bolometric luminosities
of IC and synchrotron radiation in the jet, respectively,
is plotted on the left panel of Fig.~\ref{L_bol} (maroon-solid line). 
Note that  at $z \geq 1$~pc, where $R_{\rm sp} \propto z$,  
$L_{\rm ntj} \sim 10^{37}$~erg~s$^{-1}$ is constant on $z$ as is shown in 
Fig.~\ref{L_bol} with a black-solid line.

\subsection{Leptonic emission from the wind shock}
\label{hadronic_w}

\begin{figure*}
\centering 
\includegraphics[angle=270, width=0.49\textwidth]{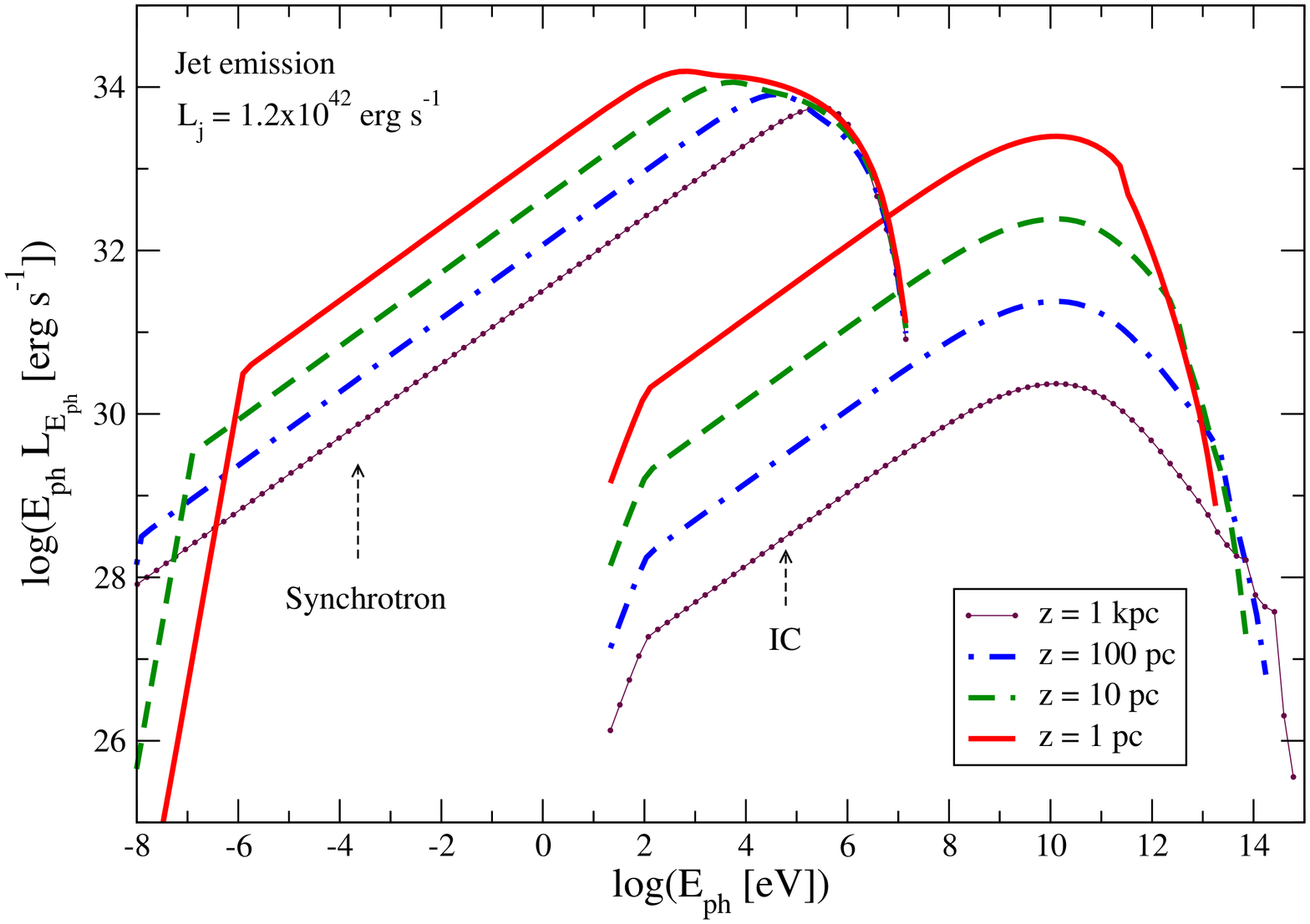}
\includegraphics[angle=270, width=0.49\textwidth]{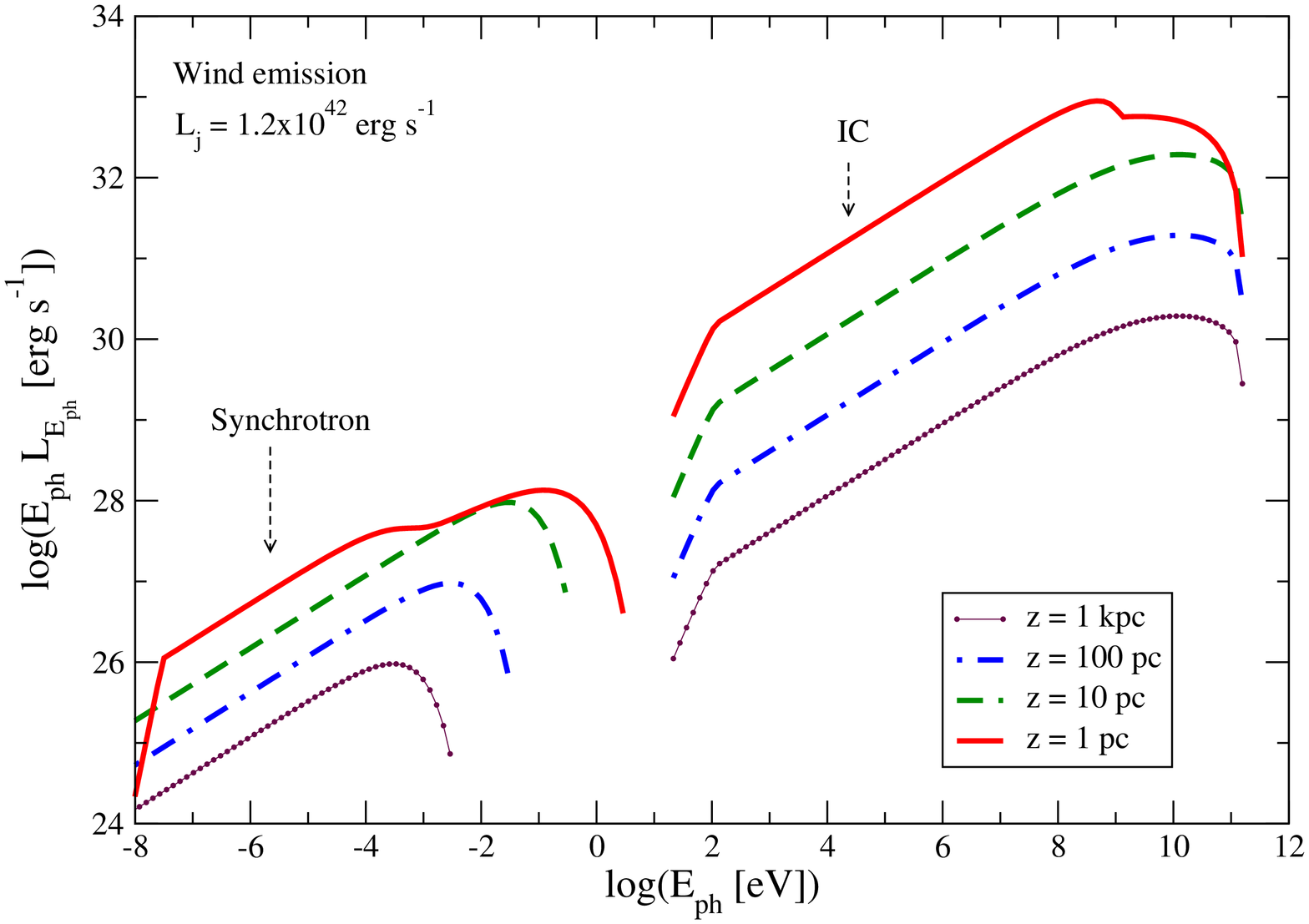}\\
\includegraphics[angle=270, width=0.49\textwidth]{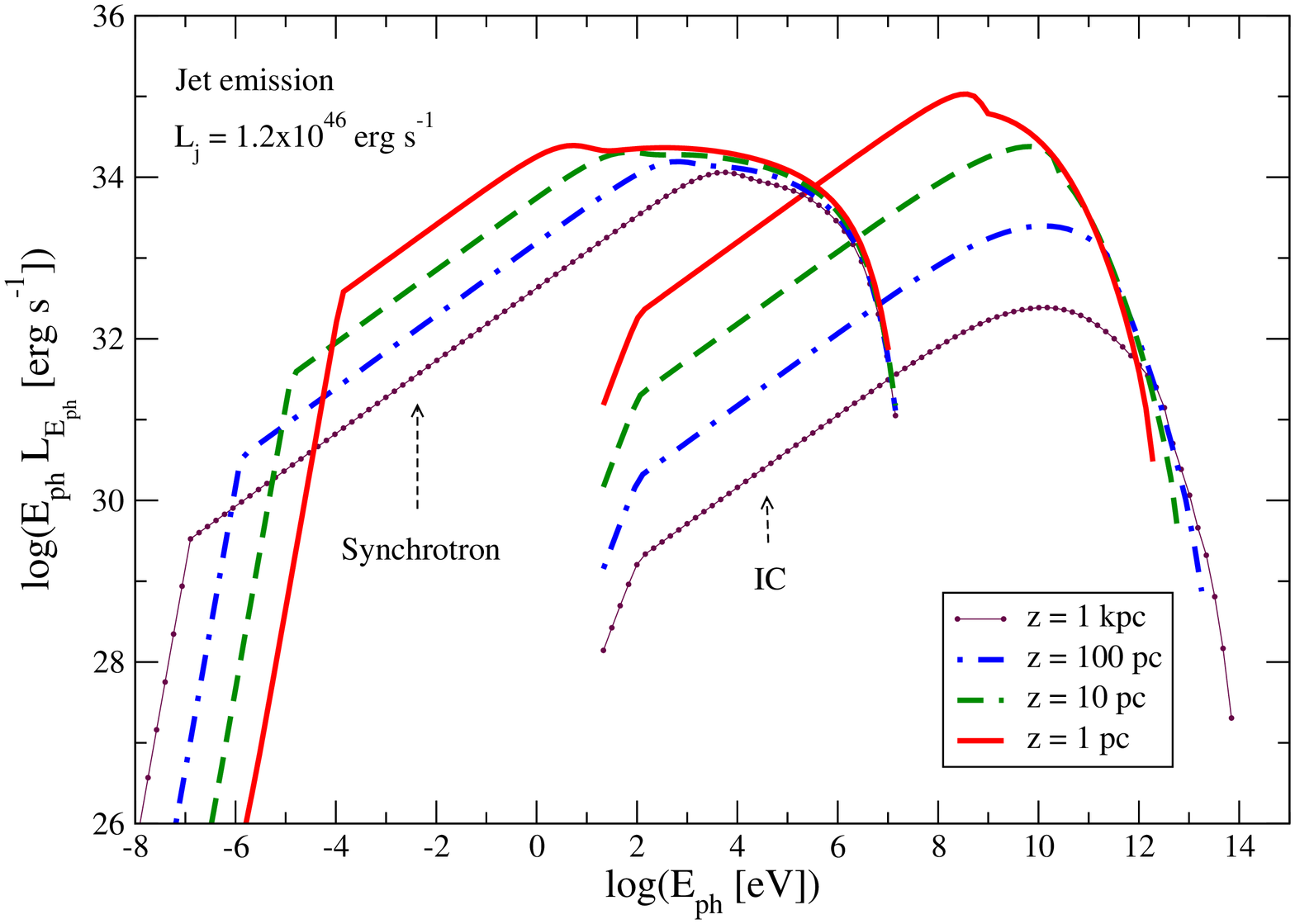}
\includegraphics[angle=270, width=0.49\textwidth]{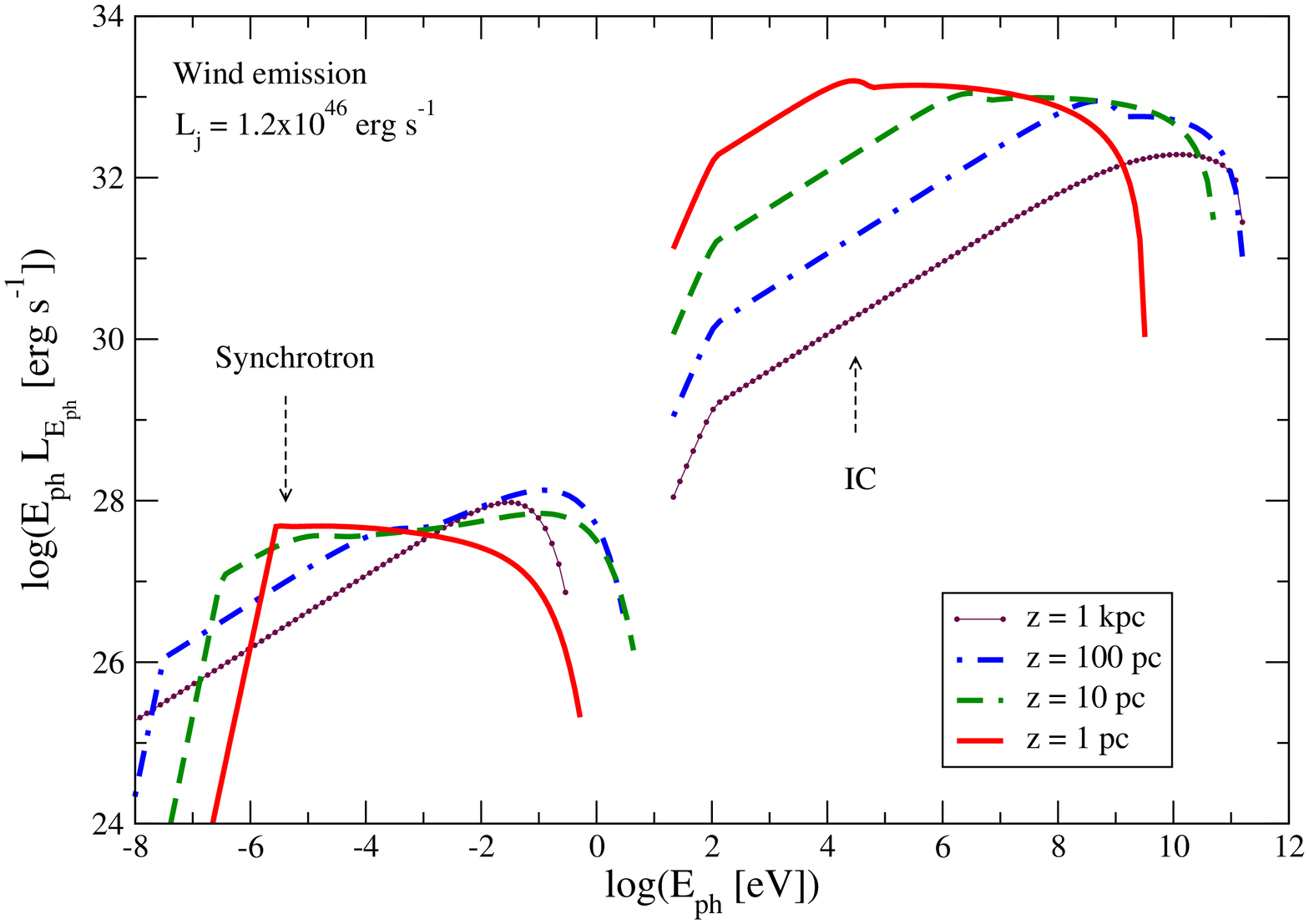}\\
\caption{Spectral energy distribution produced by the interaction of only 
one massive star with a jet of $L_{\rm j} = 1.2\times10^{42}$
(top) and  $1.2\times10^{46}$~erg~s$^{-1}$ (bottom)
at $z = 1$ (red-solid lines), 10 (green-dashed lines), 100 (blue-dot-dashed 
lines), and $10^4$~pc (maroon-dotted lines). 
The emission produced in the jet and in the
wind are  shown in the left and right panels, respectively.}
\label{sed_zfijo}
\end{figure*}

The synchrotron and IC emission from the wind bow shock is presented
in Fig.~\ref{sed_zfijo} (right panel), also for the cases of only one star 
interacting with a jet of $L_{\rm j} = 1.2\times10^{42}$ (top) and 
$1.2\times10^{46}$~erg~s$^{-1}$ (bottom),  at $z = 1$, 10, 100, and $10^4$~pc.
The SED shows lower maximum energies and lower achieved emission levels than 
those of the shocked jet region.
We can see from the figure that the synchrotron
emission produced in the wind is very faint, with an specific luminosity 
about five order of magnitude lower than the IC emission.

\begin{figure*}
\centering 
\includegraphics[angle=270,width=0.49\textwidth]{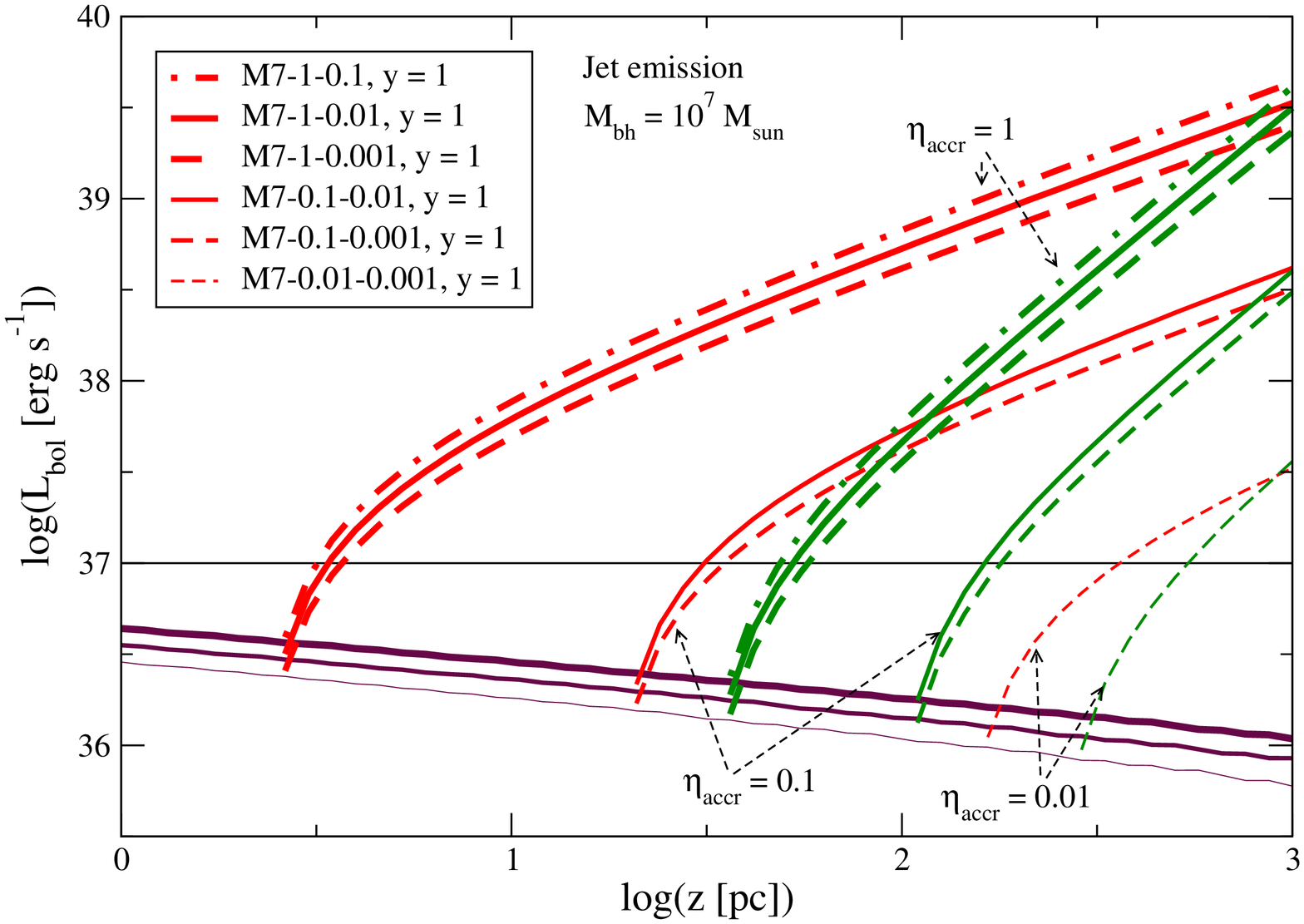}
\includegraphics[angle=270,width=0.49\textwidth]{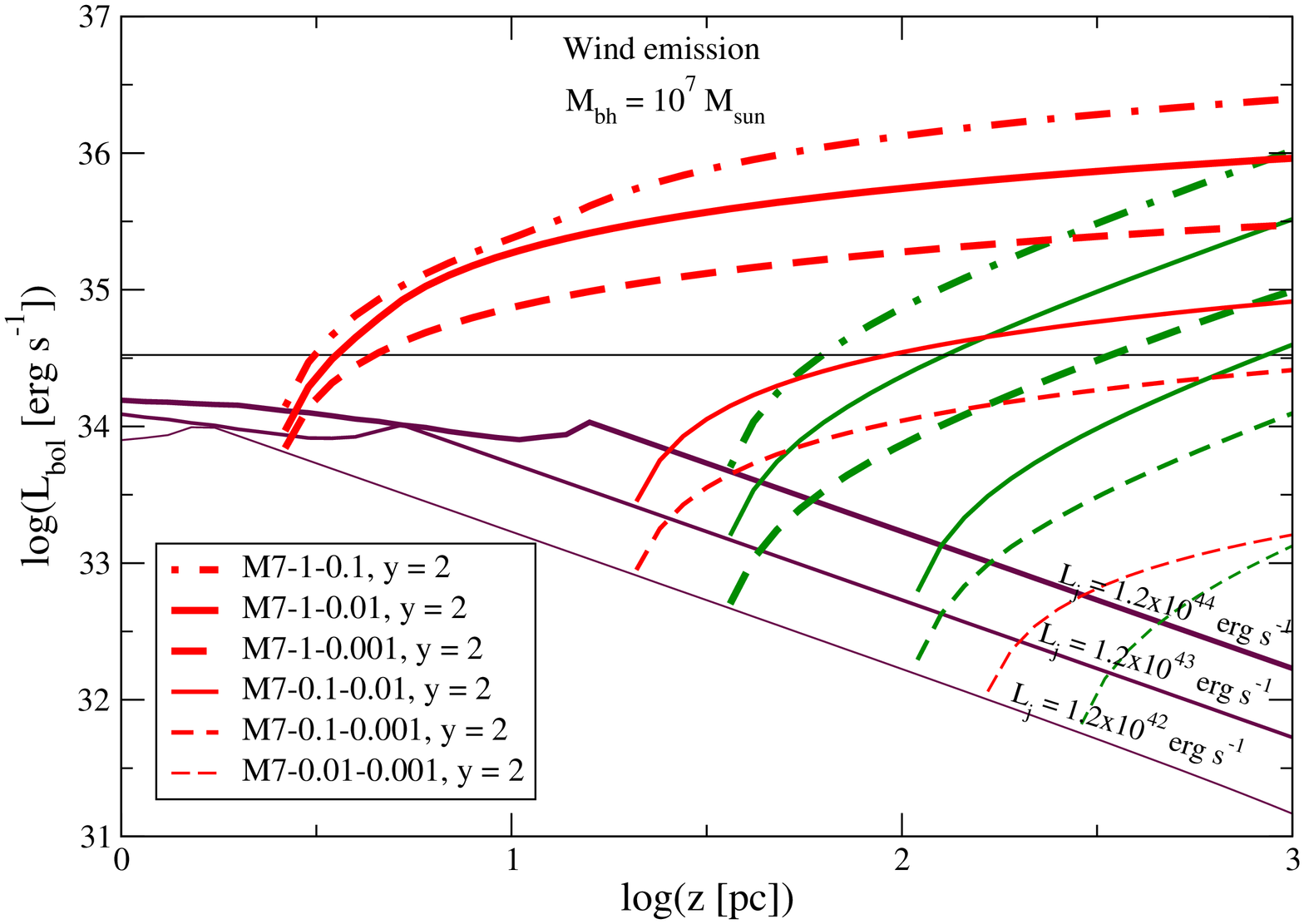}\\
\includegraphics[angle=270, width=0.49\textwidth]{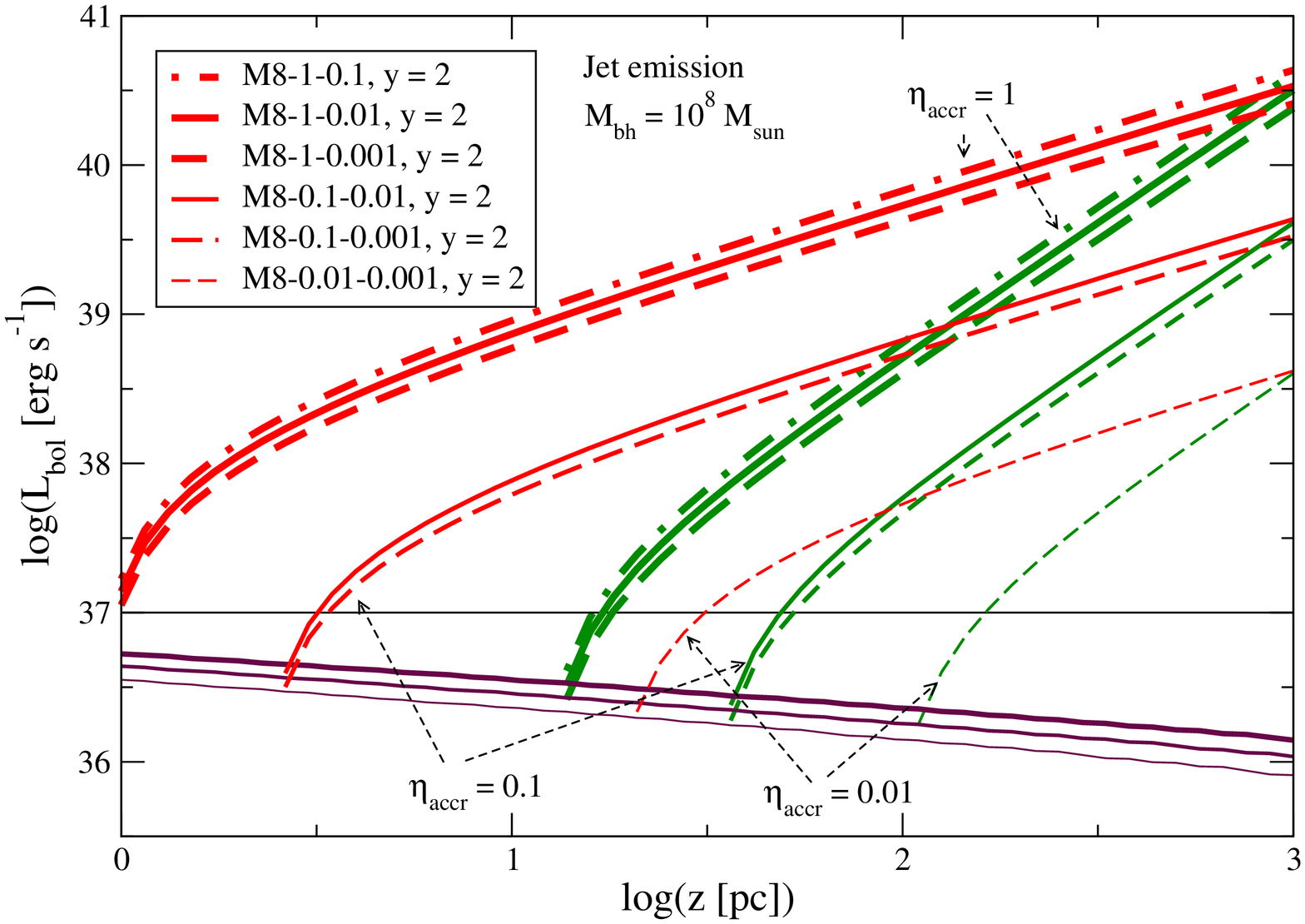}
\includegraphics[angle=270, width=0.49\textwidth]{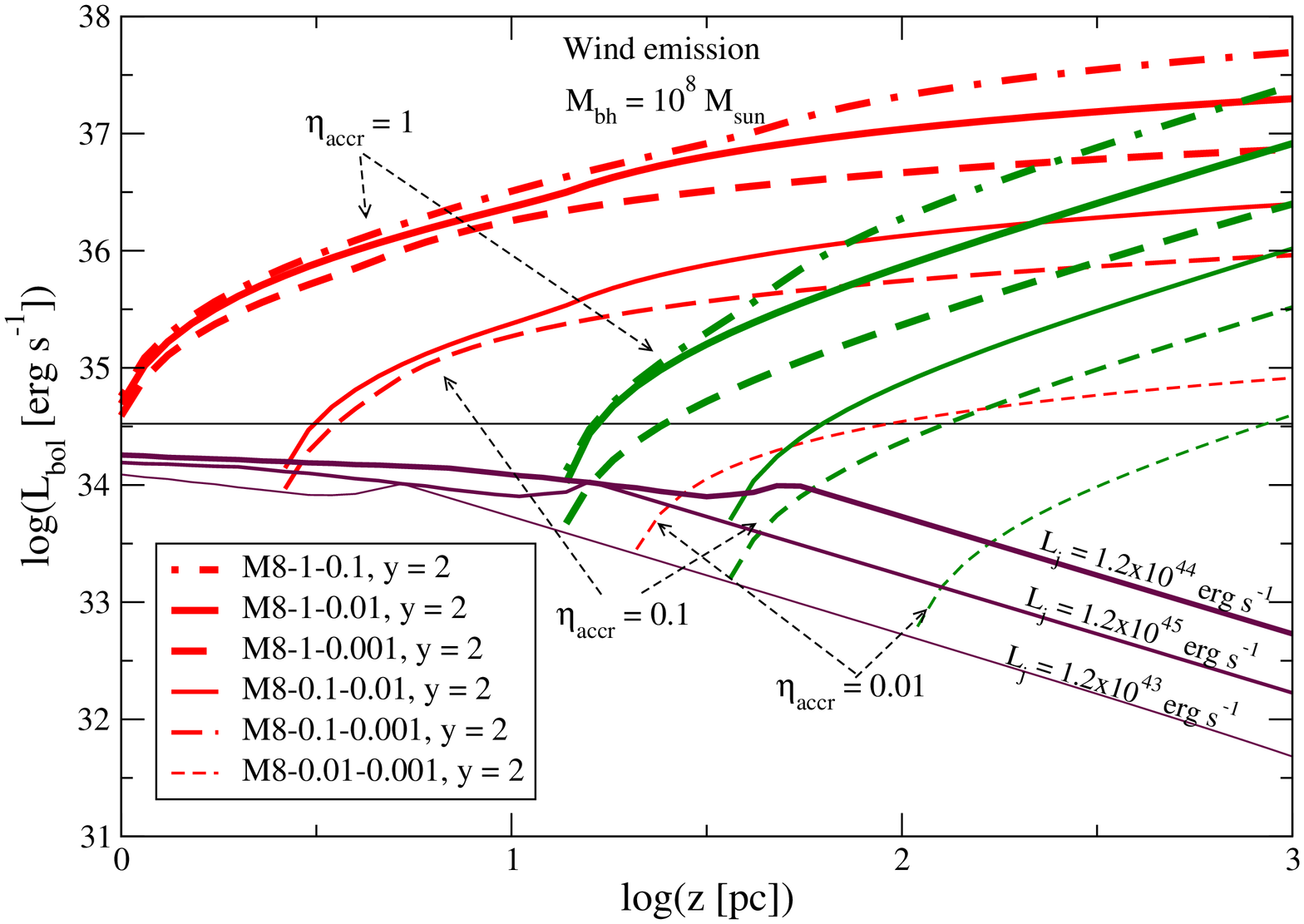}\\
\includegraphics[angle=270, width=0.49\textwidth]{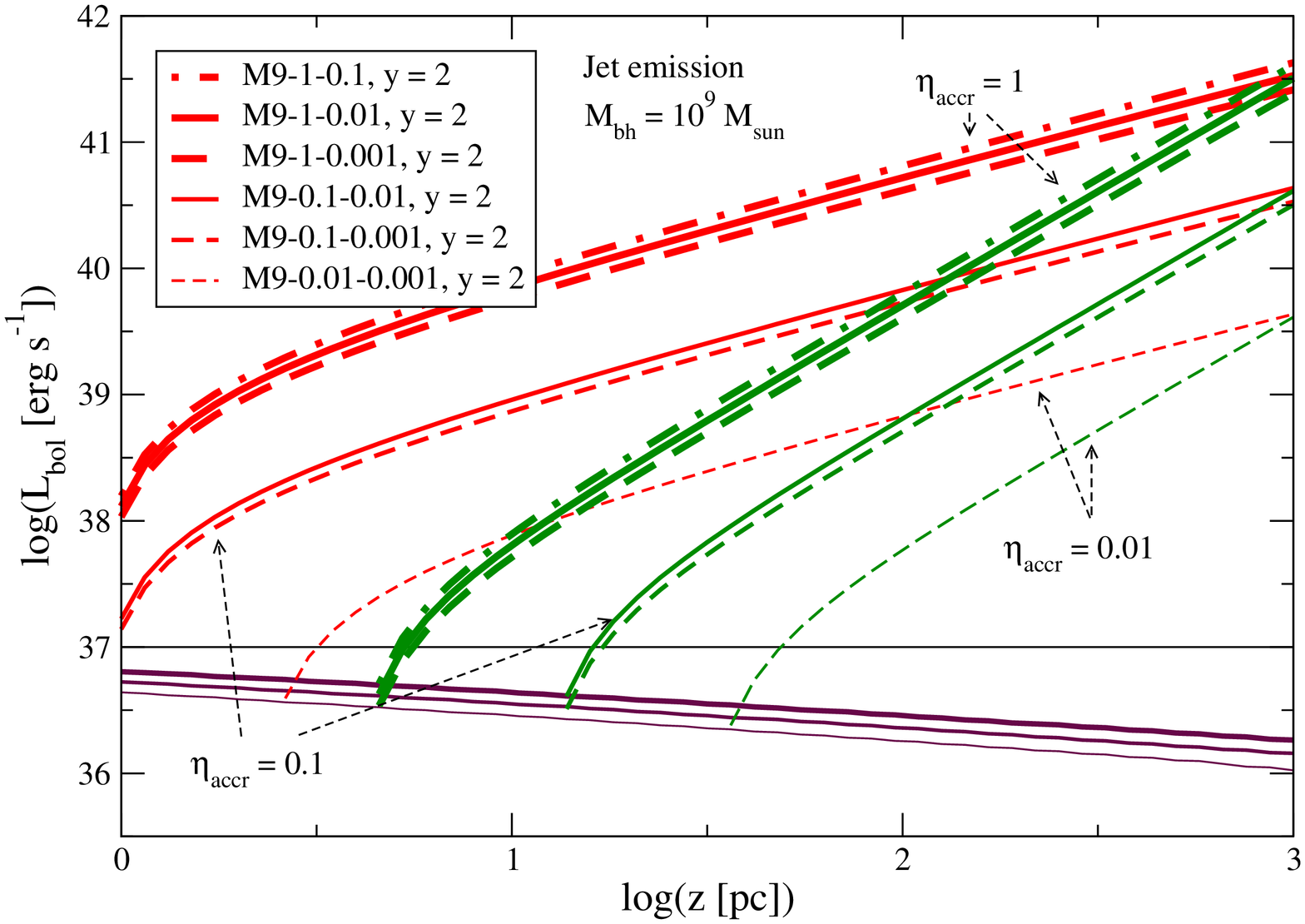}
\includegraphics[angle=270, width=0.49\textwidth]{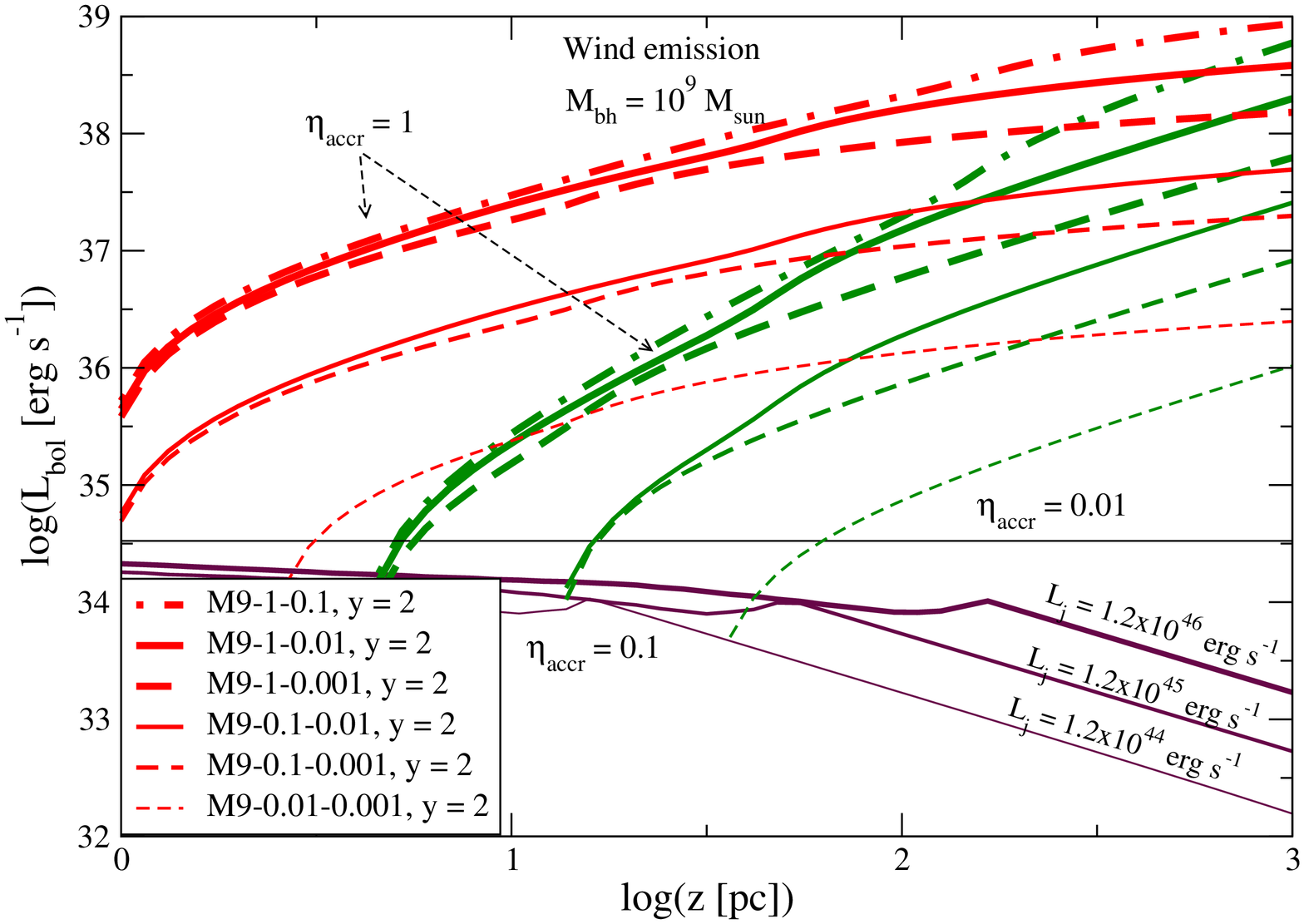}
\caption{Bolometric luminosities (maroon-solid lines) in the jet (left) 
and in the wind (right),  produced by the interaction of only one star
along the whole jet: from 1~pc to 1~kpc. The thickness of maroon lines is 
increased from low to large values of
$L_{\rm j}$, as is indicated in the right panel.
The bolometric luminosity of many stars up to a certain $z$ is also presented.
Cases with $M_{\rm bh} = 10^7$ (top), $10^8$ (middle), and $10^9$~M$_{\odot}$ 
(bottom) are shown. In each plot, the results of the two spatial distribution 
models ($y = 1$ -green lines- and 2 -red lines) are presented.   
The thickness of green and red lines is increased from low to large values of
$\eta_{\rm accr}$. The black line indicates the value of $L_{\rm jbs}$ (left), and 
$L_{\rm wbs}$ (right).}
\label{L_bol}
\end{figure*}

The total bolometric luminosity produced in the wind, 
$L_{\rm z}^{\rm w} = L_{\rm IC,z}^{\rm w} + L_{\rm synchr,z}^{\rm w} \sim  L_{\rm IC,z}^{\rm w}$,
is plotted on the right panel of Fig.~\ref{L_bol} (maroon-solid line). 
Note that at $z \geq 1$~pc, $L_{\rm z}^{\rm w} \propto  z^{-1}$.
Finally, note that as a consequence of 
$t_{\rm adv}^{\rm w}/t_{\rm adv}^{\rm j} \sim 100$, because $v_{\infty}/c \sim 100$,
the fraction of the available non thermal luminosity that is radiated in the 
wind  is larger than in the case of the jet emission, i.e. 
$L_{\rm z}^{\rm w}/L_{\rm ntw} > L_{\rm z}^{\rm j}/L_{\rm ntj}$. 

Although $n_{\rm w}$ is larger than in the shocked jet region, the production 
of gamma rays by proton-proton interactions of wind accelerated protons 
and shocked matter is
negligible when compared with emission from IC scattering. For this
reason, we do not compute the luminosity produced by this emission channel.
Besides that, the synchrotron and IC emission from $e^{\pm}$ secondaries of 
these proton-proton interactions will be much smaller than that
from primary electrons.


\section{Flaring emission from a Wolf-Rayet star}

Wolf-Rayet stars evolve from OB-type stars.  Typically, WR stars have
masses $\sim 10-25$~M$_{\odot}$, and strong mass-loss rates, $\sim
10^{-4}$~M$_{\odot}$~yr$^{-1}$. They are very luminous, $L_{\rm
WR}\sim 10^{39}$~erg~s$^{-1}$,  reaching photospheric radius as large
as  $\sim 10^2$~R$_{\odot}$ in the most powerful cases
\citep{crowther_07}.  Since WR stars are scarce, it is not expected to find 
large populations of WR stars  in the inner region of AGN, we will consider
here the situation of a single WR star interacting with the AGN
jet. The winds of WR stars are so powerful that can balance the ram
pressure of a jet with $L_{\rm j0} = 1.2\times10^{42}$~erg~s$^{-1}$ at any $z$,
since $z_{\star} \sim 0.74 \, (L_{\rm j0}/10^{42}$~erg~s$^{-1})^{1/2} z_0$ for 
the properties of the WR star listed on Table~\ref{wr_table}.

\begin{table}
\caption{Parameters of the WR star considered in this work.}
\begin{tabular}{ll}
\hline\hline
Description & Value\\
\hline
Mass loss rate & $\dot M_{\rm WR} = 10^{-4}$~M$_{\odot}$~yr$^{-1}$\\
Wind terminal velocity & $v_{\rm WR} = 3000$~km~s$^{-1}$\\
Luminosity & $L_{\rm WR} = 10^{39}$~erg~s$^{-1}$\\
Surface temperature & $T_{\rm WR} = 3\times10^4$~K\\
\hline 
\hline 
\label{wr_table}
\end{tabular}
\end{table}

\begin{figure}
\includegraphics[angle=270, width=0.49\textwidth]{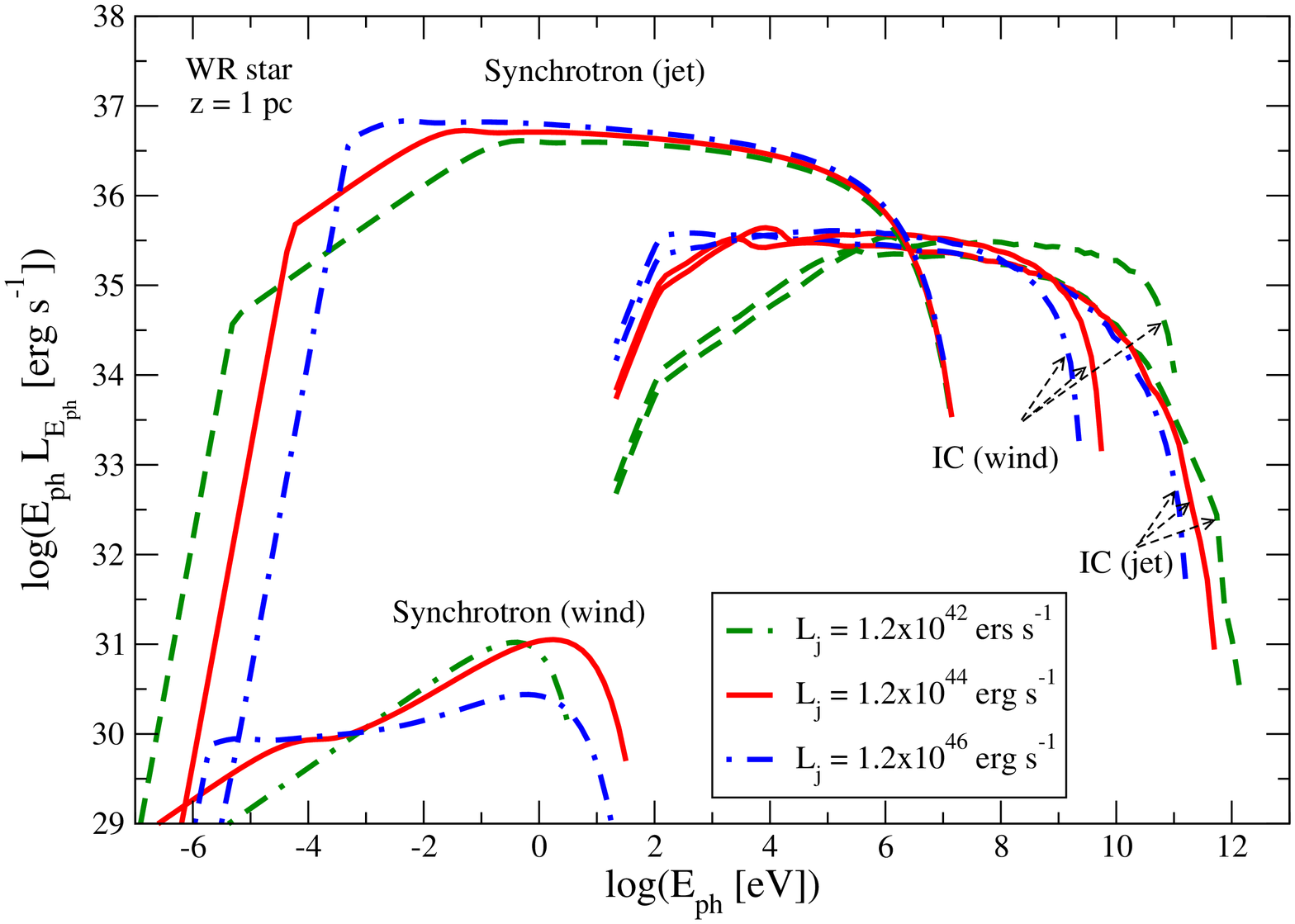}
\caption{Synchrotron radiation and IC scattering  produced in the jet 
and in the wind by the interaction of a WR star with a jet of 
$L_{\rm j0}= 1.2\times10^{42}$ (green-dashed line), $1.2\times10^{44}$ 
(red-solid line), and  $1.2\times10^{46}$~erg~s$^{-1}$ (blue-dot-dashed line) 
at $z = 1$~pc.  
The main contributions to the SED are synchrotron radiation in the jet
and IC scattering in the jet and in the wind. However, 
synchrotron emission produced in the wind is also 
plotted in order to compare this figure with Fig~\ref{sed_zfijo}.}
\label{sed_wr_z1}
\end{figure}

In order to compare the spectrum produced by a WR star and by standard 
massive (OB) stars as was shown in Section~\ref{non-thermal}, 
we assume that the WR penetrates the jet at $z = 1$~pc in the case 
with $L_{\rm j0} = 1.2\times10^{42}$, $1.2\times10^{44}$, and  
$1.2\times10^{46}$~erg~s$^{-1}$.
Being $\dot M_{\rm WR}/\dot M_{\star} = 100$, the stagnation point of the
WR wind is located at $R_{\rm sp,wr} \sim 10 R_{\rm sp}$. Thus,
the available luminosity to accelerate particles in the shocks produced
by the interaction of the WR is $\sim 100$ times larger than in the case
of an OB star. In Fig.~\ref{sed_wr_z1}
we show the synchrotron and IC emission  produced in the jet 
and in the wind. Note  that the IC emission from the wind reaches similar 
levels to the IC emission from the jet, on the contrary to the case of an
OB star, where the IC jet emission in the case of 
$1.2\times10^{46}$~erg~s$^{-1}$ is $\sim 100$ times smaller in the wind
than in the jet. This is a consequence of the different energy breaks in
the electrons energy distribution. 
Comparing the curves that correspond to $z = 1$~pc in 
Fig.~\ref{sed_zfijo} with Fig.~\ref{sed_wr_z1} we can appreciate that the 
shape of synchrotron and IC spectrum in the jet is  different for the case 
of an OB star and a WR, where in the former case the break energy produced 
by the advenction escape to the radiation dominated regime is at higher 
energies than in the latter.  
Finally, the emission level produced by a WR (both in the wind
and in the jet) is larger 
than the one produced by an OB star (both interacting with the jet at the 
same $z$).

The radiation
produced by a WR interacting occasionally with a jet will be
transient with a timescale similar to the jet crossing time, 
unlike the steady emission produced by a population of
stars, described in the next section. 
We remark that, if the star diffusion time were short enough to allow a
massive star to reach the
vicinity of the SMBH in the WR stage, the luminosity due to the jet-WR
interaction would be significantly higher than obtain for an
interaction distance of 1 pc.
It is noteworthy that
one or few WR may be recurrently present within the jet and close 
to its base, where radiative cooling is still dominant, adding up to
the contribution of the many-star persistent emission. In fact, WR
could be important contributors of their own to the non-thermal output
of misaligned AGN jets.


\section{Steady emission from a population of massive stars}
\label{Many_stars}

\begin{figure*}
\centering 
\includegraphics[angle=270, width=0.49\textwidth]{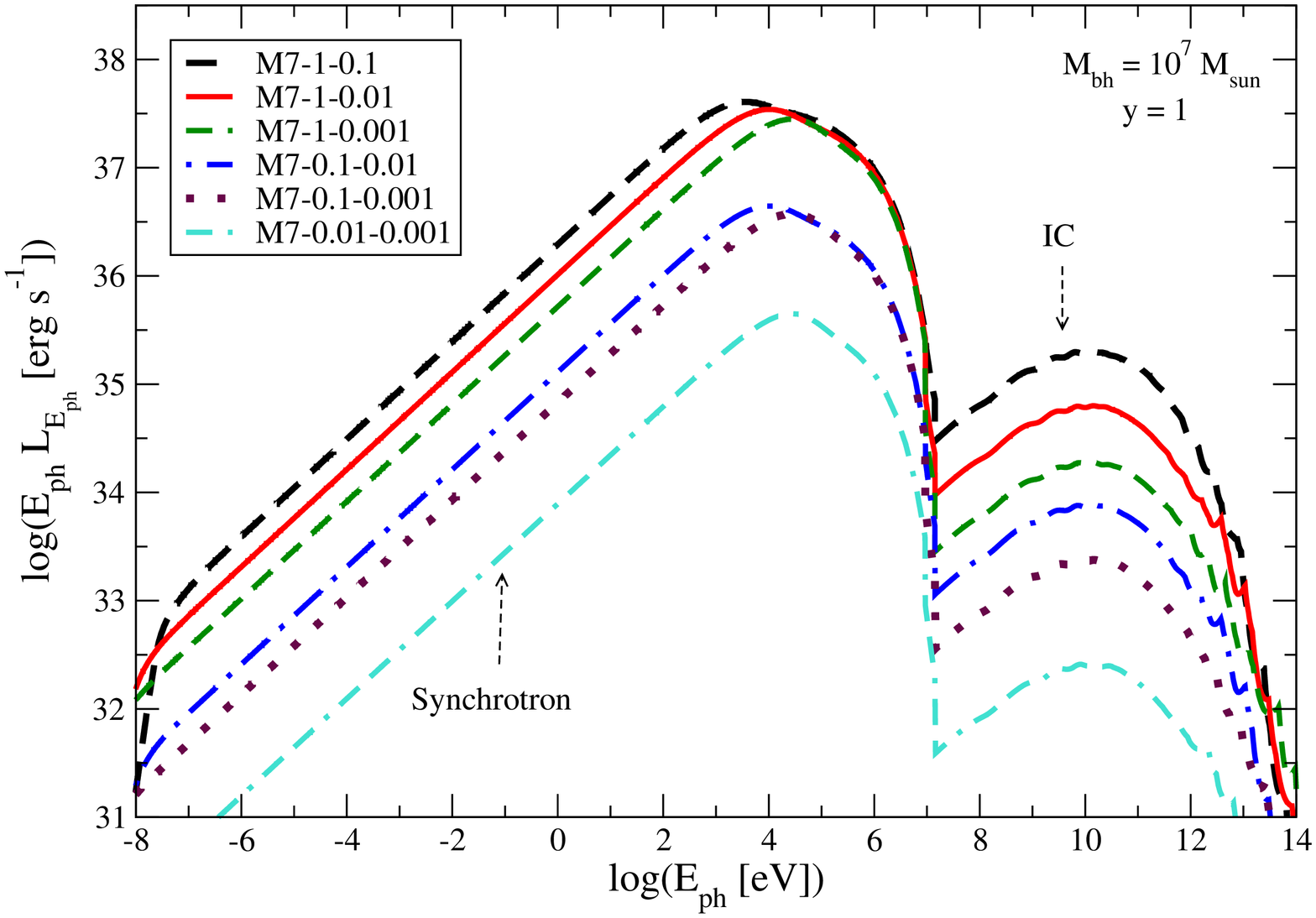}
\includegraphics[angle=270, width=0.49\textwidth]{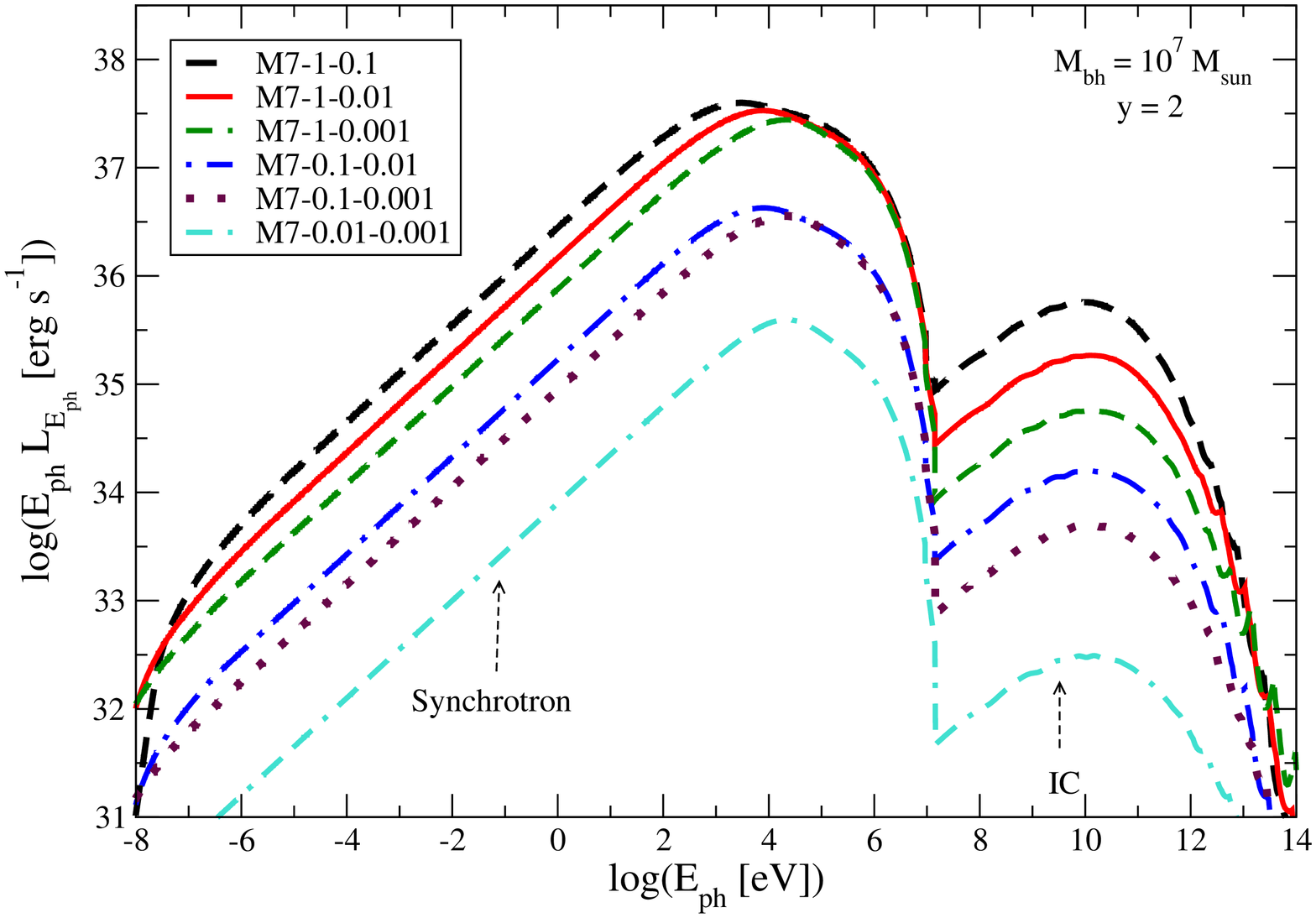}\\
\includegraphics[angle=270, width=0.49\textwidth]{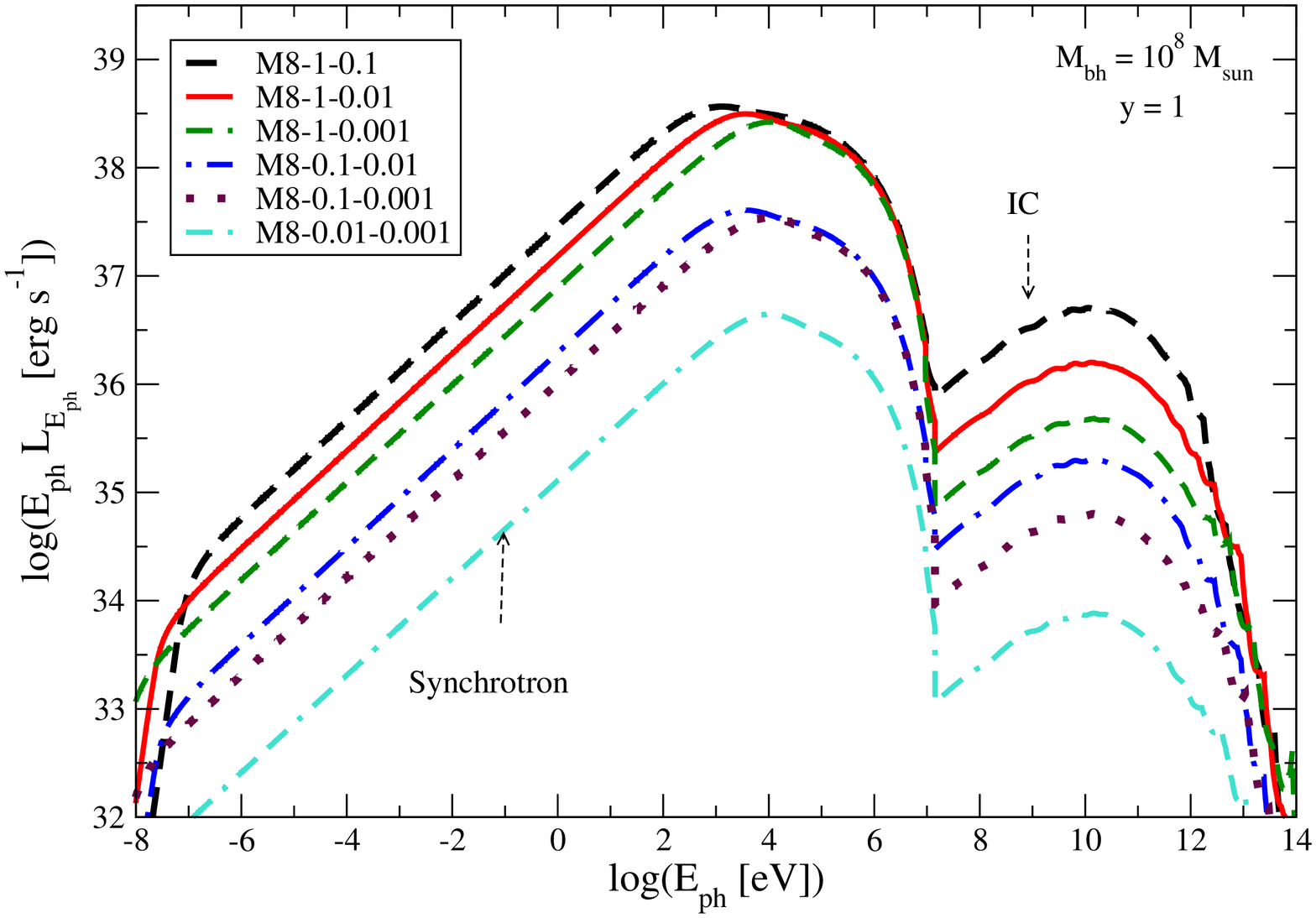}
\includegraphics[angle=270, width=0.49\textwidth]{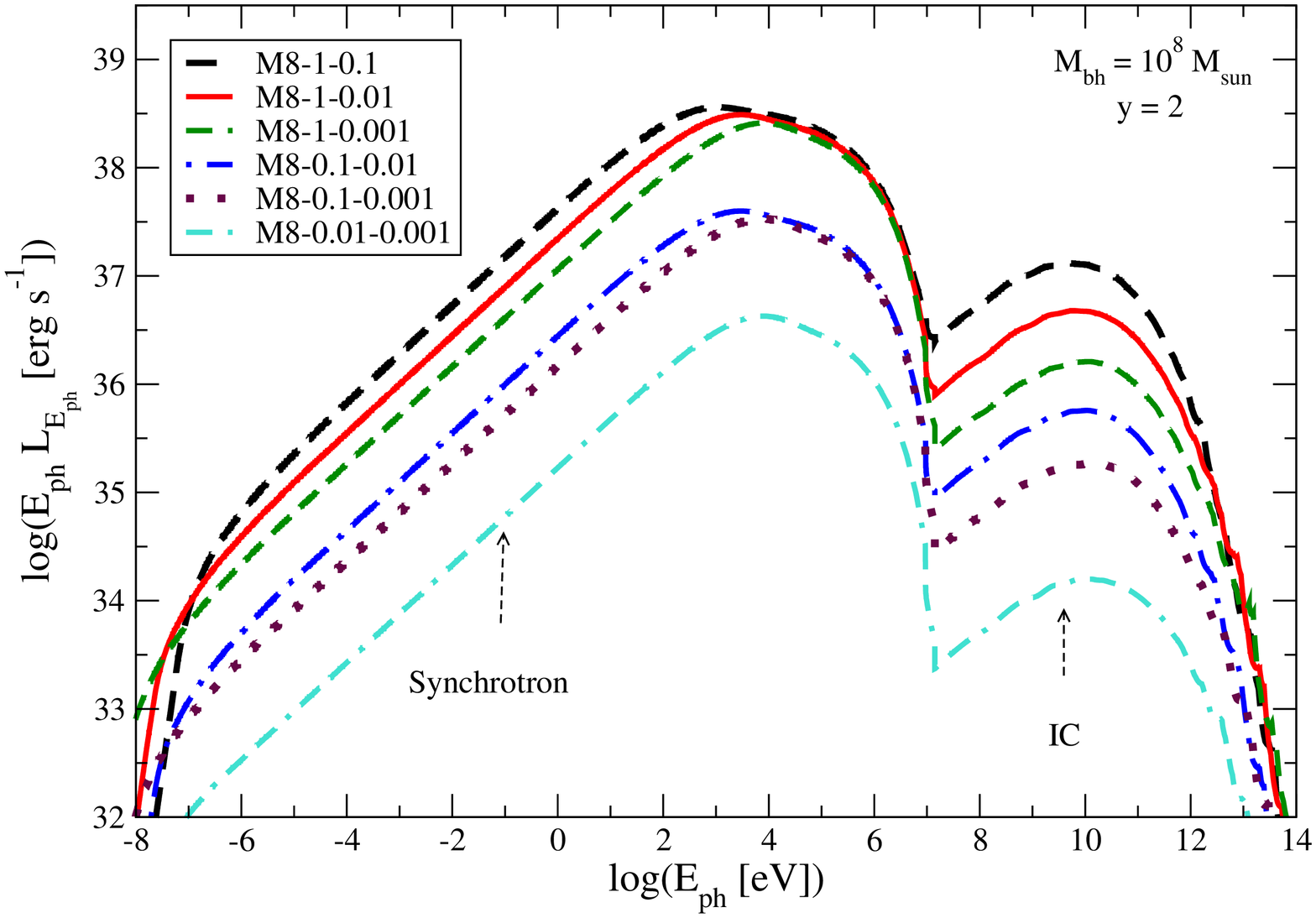}\\
\includegraphics[angle=270, width=0.49\textwidth]{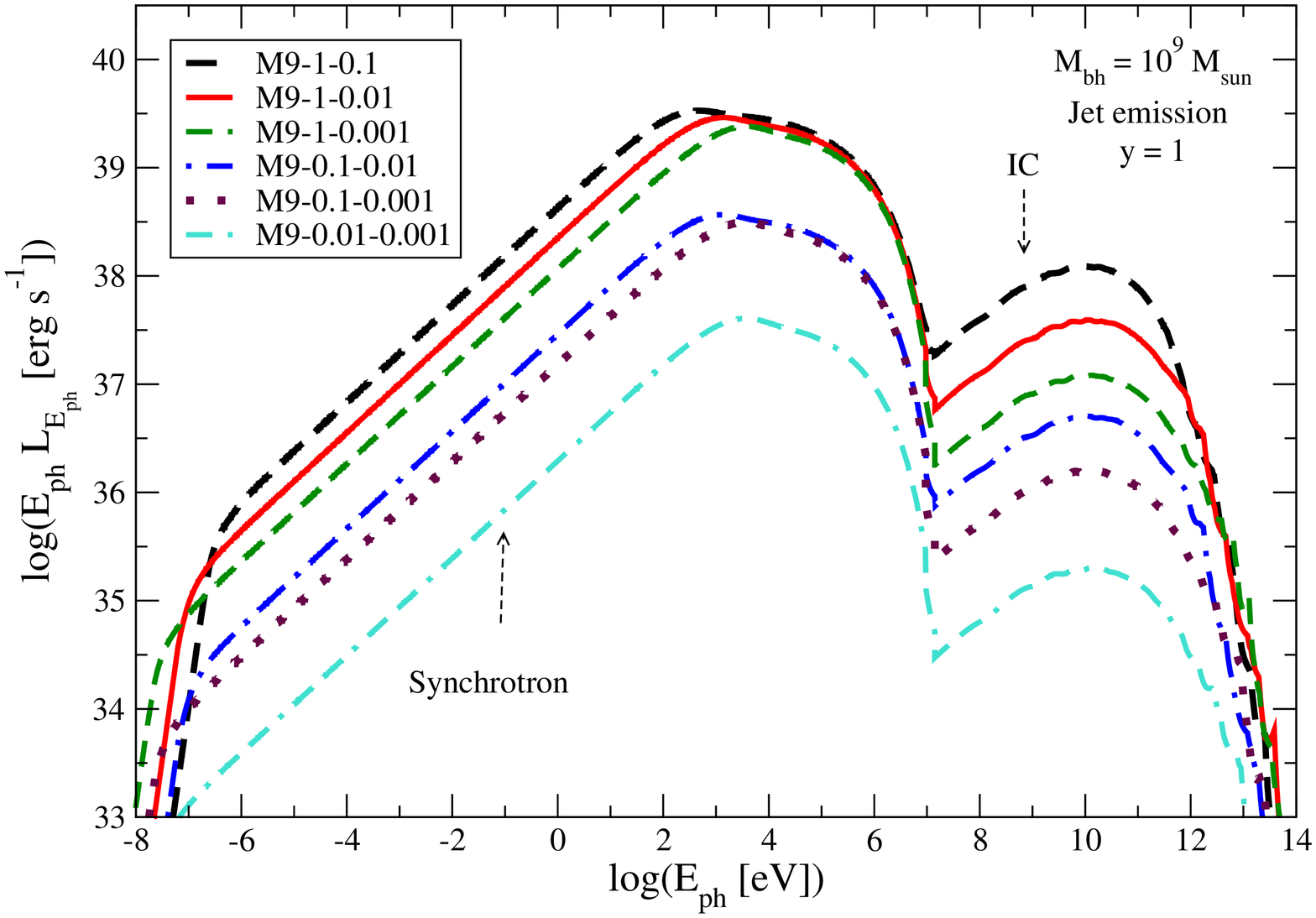}
\includegraphics[angle=270, width=0.49\textwidth]{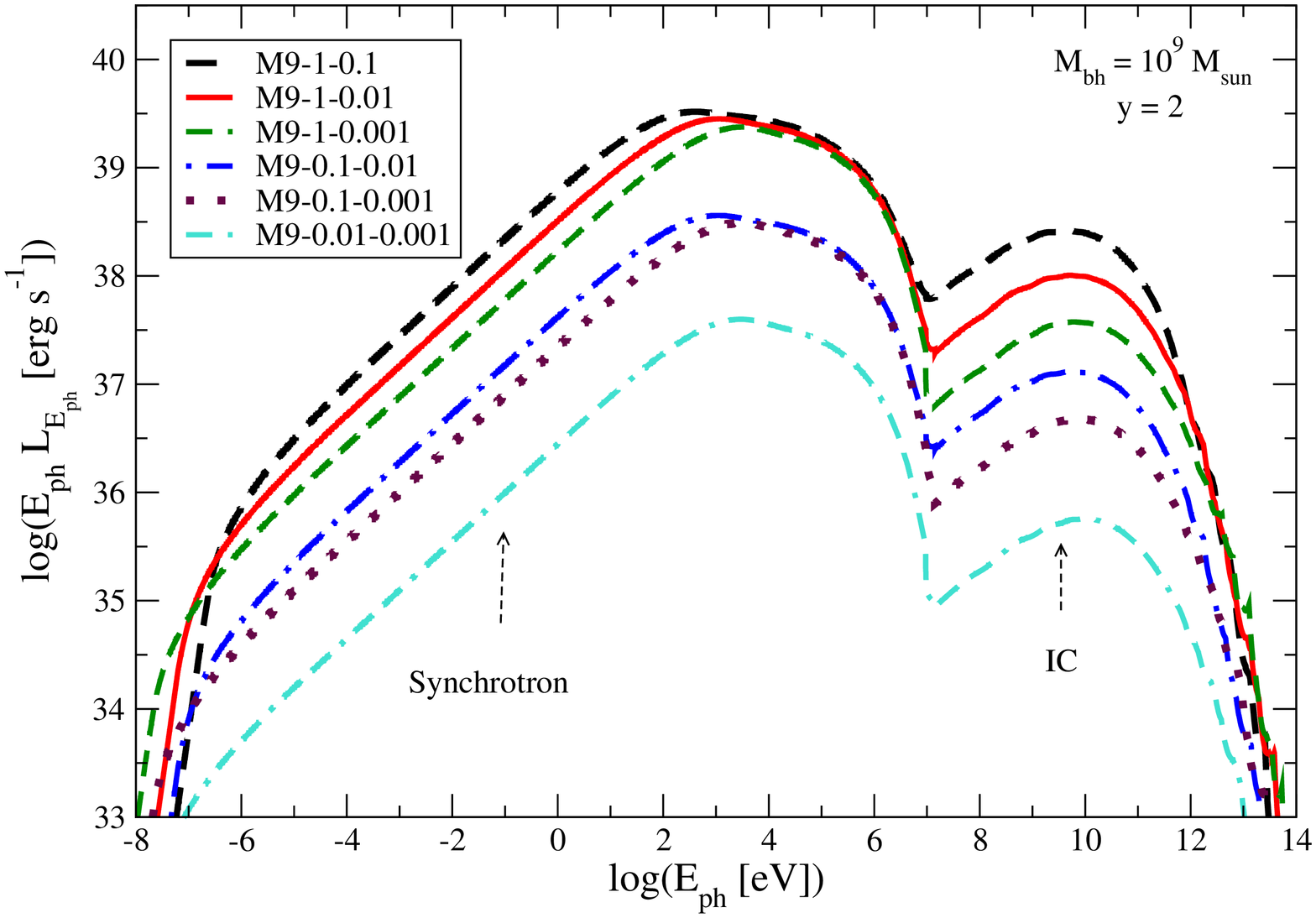}
\caption{Spectral energy distribution of the emission up to $z =$~1kpc 
produced by $N_{\star \rm j}$ stars inside the jet.  
The main contributions to the SED are synchrotron radiation 
and IC scattering; proton-proton interactions are not relevant. 
Left panel is for the case of $y = 1$, and the right one for $y = 2$.   
Cases with $M_{\rm bh} = 10^7$ (top), $10^8$ (middle), and $10^9$~M$_{\odot}$ 
(bottom) are shown.}
\label{sed_stars}
\end{figure*}

In order to study the emission produced by many massive stars,  we
assume within the jet a stellar population as the one described in
Sect.~\ref{sec_population}.  As shown in Sect.~\ref{non-thermal},  the
emission produced at small values of $z$ is higher than the emission
produced at larger $z$, as a consequence of the dilution of the target
fields with $z$. This effect is balanced by the fact that, at $z >
z_1$, the number of stars interacting with the jet is $> 1$ and the
emission produced by all of them increase $\propto z^2$ and $\propto z$,
for the cases with $y = 1$ and 2, respectively. We
calculate the emission produced by each of the $N_{\star \rm j}$
stars at a certain $z$, and then integrate along $z$ all the
contributions, obtaining the SEDs shown in Fig.~\ref{sed_stars},
for different values of $M_{\rm bh}$ and $L_{\rm j}$.  Note that the
features of these SEDs are similar to the SED produced by only one
star located at a relatively large value of $z$ (see Fig.~\ref{sed_zfijo}), 
where advection losses become dominant.  In the
range $z > 1$~pc, $R_{\rm sp}$ is large enough  to suppress the
effect of photon-photon absorption. In the case of $L_{\rm j} =
1.2\times10^{46}$~erg~s$^{-1}$, the synchrotron and IC
emission achieve levels of $\gtrsim 5\times 10^{39}$~erg~s$^{-1}$ in 
hard X-rays and $\sim 10^{38}$~erg~s$^{-1}$ in gamma rays, respectively.

In Fig.~\ref{L_bol}, the bolometric luminosities (synchrotron + IC)
at different $z$ and for a variety of stellar distributions are
shown. In Section~\ref{non-thermal} we have commented about 
the non-thermal bolometric luminosity ($L_{z}^{\rm j}$ and $L_{z}^{\rm w}$) 
produced by the interaction of only one star with the jet at different $z$,
from 1~pc to 1~kpc (maroon-solid lines). However, at $z \gtrsim z_1$ there 
are more
than one star every time into the jet, and the  non-thermal luminosity
produced by all the stars into the jet at different $z$ is also plotted
in  Fig.~\ref{L_bol}. In each panel we show the bolometric luminosity 
produced by the different stellar populations considered in the present study.  
Note that on the one hand, in the most powerful case (M9-1-0.1) 
the total bolometric
luminosity produced in the jet and in the wind is $\sim 5\times10^{41}$ and
$\sim 10^{39}$~erg~s$^{-1}$ ($y = 2$), respectively. On the other hand, 
in cases with low density of massive stars, 
the luminosity produced by the cumulative effect of all stars into the
jet can be lower
than the luminosity produced by only one star interacting with the jet
close to $z_0$ if the star
formed at $z \gtrsim r_{\rm t}$ and migrated close to the jet base.

Considered the $\dot M_{\star}-\dot M_{\rm bh}$ relation given by 
\cite{Satypal_05}, the density of massive stars results 
$\propto (\eta_{\rm accr}\,M_{\rm bh})^{0.89}$. Thus, sources with 
$M_{\rm bh} = 10^8 - 10^9$~M$_{\odot}$ and $\eta_{\rm accr} \sim 1 - 0.1$
are likely to be detected at gamma rays by \emph{Fermi}  with a 
deep enough (pointed) exposure or after some years of observation in the 
survey mode. In the case of stellar populations around a SMBH with 
$M_{\rm bh} = 10^7$~M$_{\odot}$, the gamma ray emission produced cannot be
detected by  \emph{Fermi} in any case, and the same occurs for 
$M_{\rm bh} = 10^8$~M$_{\odot}$ and $\eta_{\rm accr} \sim 0.01$ 
(under the assumed $\eta_{\rm B}$ and $\eta_{\rm nt}$). 
The most interesting case is that of a high accretion rate 
$\eta_{\rm accr} \sim 1$ and
$M_{\rm bh} = 10^9$~M$_{\odot}$~yr$^{-1}$, whose emission can be detected  
in the case of luminous  ($L_{\rm j} \sim 10^{46}$~erg~s$^{-1}$) and 
close sources (such as M87).
Less luminous sources may also be detected in the near future by 
the  Cherenkov Telescope Array (CTA).    

In the case of a population of massive stars (continuously)
interacting  with the jet, the  produced emission will be steady and
produced in a large part of the jet volume, from $z_{\rm 1}$ to
$z_{\rm 2}$, on scales of $\sim$~kpc.


\section{Discussion and summary}

In this work we have studied the interaction of massive stars with
relativistic jets of AGN, focusing on the production of gamma rays
from particles accelerated in the double bow-shock structure formed
around the stars as a consequence of the jet/stellar wind
interaction. We calculated the energy distribution of electrons 
accelerated in the jet and in the wind, and the
subsequent non-thermal emission from these relativistic particles. In
the jet and wind shocked regions, the most relevant radiative processes are
synchrotron emission and IC scattering of stellar photons. In the
wind shocked region, the gamma-ray luminosity from proton-proton interactions
in the stellar wind is well below the IC one. 

We have studied two scenarios: the interaction of a WR
star {\bf at 1~pc}; and the interaction of a population of
massive stars with the whole jet.  The properties of
the emission generated in the downstream region of the bow shocks
change with $z$. On the one hand, the target densities for radiative
interactions decrease as $z^{-2}$.  On the other hand, the time of the
non-thermal particles inside the emitter is $\propto R_{\rm sp}\propto z$, 
and the number of stars per jet length unit 
${\rm d}N_{\star \rm,j}/{\rm d}z\propto z$ 
and $z^{2}$, for cases with $y = 2$ and 1, respectively. 
Therefore, for a population of stars, the last two 
effects soften the emission drop with $z$.

The interaction of only one star with the jet can produce significant amounts of
high-energy emission only if the interaction height is below the $z$
at which advection escape dominates the whole particle
population. Also, $\sigma_{\rm sp}$ should be a significant fraction
of $\sigma_{\rm j}$.  In this context, we have considered the
interaction of a powerful WR star at $z = 1$~pc. The
emission produced by IC scattering achieves values as
high as  $\gtrsim 10^{36}$~erg~s$^{-1}$ 
(considering the contribution of the wind and 
jet in Fig.~\ref{sed_wr_z1}) in the  \emph{Fermi} range. Such an
event would not last long though, about $R_{\rm j}/v_\star\sim
300\,(R_{\rm j}/3\times10^{17}\,{\rm cm})\,(10^9\,{\rm
cm~s}^{-1}/v_\star)^{-1}$~yr. The emission level could be detectable
by \emph{Fermi} only for very nearby sources, like Centaurus~A (located
at a distance $d \sim 4$~Mpc).  The interaction of few WR stars interacting
with jets in more distant sources like M87 ($d \sim 16$~Mpc) could be
detectable by the forthcoming CTA.
The interaction of a star even more powerful than a WR, like a
Luminous Blue Variable, may provide $R_{\rm sp}\sim R_{\rm j}$, making
available the whole jet luminosity budget for particle acceleration.

In the middle/end part of the jet, the interaction of many massive
stars  can also produce a significant amount of gamma
rays. The resulting SED integrated along the whole jet strongly
depends on the number of stars inside it. We have considered a
Salpeter initial mass function of stars distributed following a power-law 
spatial distribution
(Eq.~(\ref{psi_0})). In the case of $M_{\rm bh} = 10^{9}$~M$_{\odot}$,
and high accretion rates ($\eta_{\rm accr} = 1$),
gamma-ray luminosities $\sim  10^{38}$ and 
$5\times10^{38}$~erg~s$^{-1}$, for $y=1$ and 2, respectively, may be
achieved (see Fig.~\ref{sed_stars}). However, note that few WR 
inside the jet 
could actually dominate over the whole main-sequence OB star population.

Although  jet/star interactions are very sporadic near the base of the
jet, we note that at $z < 1$~pc, clouds from the BLR can also interact
with the jet, leading to significant gamma-ray radiation (Araudo et
al. 2010). The produced emission in BLR clouds interacting with jets
has a stronger dependence on $L_{\rm j}$ than in the case of stellar
winds, because clouds do not have winds and their cross section does
not get adjusted to ram pressure balance. Thus, jet/BLR cloud interactions
could be more relevant in sources like FR~II~galaxies.    

An interesting (similar) scenario is the interaction of a star forming
region (SFR) with the jet. There is evidence that SFRs are located in
the torus of some AGN (starburst galaxies), at distances $\sim 100$~pc
from the nucleus. In addition, hints of SFRs located in the nuclear
region of AGN are also found in galaxies with IR nuclear excess. These
galaxies are called nuclear starburst galaxies. The number of OB-type
stars in SFRs can be as high as $\sim 10^4$, distributed in a small
volume of $\sim 10$~pc$^3$. Then, if one of these compact SFRs
interact with the jet at $z \sim 10$~pc, the total luminosity could
reach detectable levels, with the resulting radiation presenting rich
and complex features. Furthermore, the jet passing through the
intra-cloud rich medium can have interesting consequences in the SFR
evolution. This scenario will be analyzed in detail in a following paper.     

It is noteworthy that, for $\eta_{\rm accr} \lesssim 1$, one expects $\sim 10^4$
massive stars up to $\sim 1$~kpc. Moreover, as shown in
Sect.~\ref{clumps}, the shocked stellar wind will efficienctly mix
with the jet. Assuming an average $\dot{M}_{\rm w}\sim
10^{-6}\,$M$_\odot$~yr$^{-1}$, one can estimate the power required to
accelerate this mass to the jet Lorentz factor, $\Gamma\,\dot{M}_{\rm
w}\,c^2\sim 6\times 10^{44}$~erg~s$^{-1}$. Despite this is just an
order of magnitude estimate, this power tells us that the dynamics of
jets with similar or smaller power, i.e. $\lesssim
10^{45}$~erg~s$^{-1}$, will be significantly affected by wind
mass-loading \citep[e.g.][]{Hubbard_06}. Therefore, early-type stars,
as low-mass ones \citep{Komissarov_94}, cannot be neglected when
studying jet propagation and evolution in galaxies with moderately
high star formation. Even for $\eta_{\rm accr} \sim 0.01$, jets with $L_{\rm
j}\sim 10^{42}$~erg~s$^{-1}$ may be strongly affected by the
entrainment of wind material (see also the discussion on
mass-load in \cite{Valen-Manel-Maxim}). 

Finally, we remark that since jet-star emission  should be rather
isotropic (as in all the cases of jet-obstacle interactions), it would
be masked by jet beamed emission in blazar sources. Misaligned
sources however do not display significant beaming, and for those
cases jet-star interactions may be a dominant gamma-ray production
mechanism. In the context of AGN unification \citep[e.g.][]{Urry-Padovani},
the number of non-blazar AGN should be much larger than that of
blazars with the same $L_{\rm j}$. Close and powerful sources could be
detectable by deep enough observations of the \emph{Fermi} gamma-ray
satellite. After few-year exposure, a significant signal from
jet-star interactions could be found, and their detection would shed light 
not only on the jet properties but also on the stellar populations in the
vicinity of AGN. The same applies to stars with powerful winds
penetrating the jet at its innermost regions, which may be seen as
occasional, transient month-scale gamma-ray events.


\section*{Acknowledgments}
The authors thank the referee for valuable comments that improved
the paper. We thank G. Bruzual, J. Cant\'o, S. Cellone, 
A. Raga, and L. F. Rodr{\'\i}guez for
contributing to this work with useful suggestions and comments.
A.T.A. thanks M. Orellana and P. Santamar{\'\i}a for help with numerical 
calculations. A.T.A. is very grateful for the
hospitality of the Dublin Institute of Advanced Studies (DIAS) where
this project started. 
This project is financially supported by  CONACyT, Mexico and PAPIIT, UNAM;
PIP 0078/2010 from CONICET,
and PICT 848/2007 of Agencia de Promoci\'on Cient\'{\i}fica y T\'ecnica, 
Argentina.
G.E.R. and V.B-R. acknowledge support by the former Ministerio de 
Ciencia e Innovaci\'on  (Spain) under grant AYA 2010-21782-C03-01. 
V.B-R. also acknowledges support from grant FPA2010-22056-C06-02.
\label{lastpage}

\bibliographystyle{mn2e}
\bibliography{biblio_jet-star.bib}
\end{document}